%% file: 00_main.tex
\documentclass[sigconf]{acmart}

\usepackage{booktabs} % For formal tables

\usepackage{dblfloatfix}    % To enable figures at the bottom of page

\usepackage{framed}
\usepackage{hyperref}
\usepackage{gensymb}
\usepackage{subcaption}
\usepackage{tabularx}
\newcolumntype{Y}{>{\raggedright\arraybackslash}X}
\graphicspath{ {imgs/} }

\hypersetup{draft}

\usepackage[ruled]{algorithm2e} % For algorithms

\SetAlFnt{\small}
\SetAlCapFnt{\small}
\SetAlCapNameFnt{\small}
\SetAlCapHSkip{0pt}
\IncMargin{-\parindent}

\usepackage{amsthm} %for definition/theorem in DP intro section

\def\plaintitle{SacCalib: Reducing Calibration Distortion for Stationary Eye Trackers Using Saccadic Eye Movements}
\def\plainkeywords{Eye Tracking; Implicit Calibration; Eye Movements; Saccades}

 \title[SacCalib: Reducing Calibration Distortion for Stationary Eye Trackers Using Saccadic Eye Movements]{\plaintitle}

\definecolor{linkColor}{RGB}{6,125,233}

\setcopyright{none}
\usepackage{changes}
\colorlet{Changes@Color}{red}% I change the color to be all red

\begin{document}
% \sloppy

% \toappearbox{\large Submitted for review to CHI'19.\newline Please do not cite or circulate.}

\title{\plaintitle}

\author{Michael Xuelin Huang}
\affiliation{%
  \department{Max Planck Institute for Informatics}
  \institution{Saarland Informatics Campus}}
\email{mhuang@mpi-inf.mpg.de}

\author{Andreas Bulling}
\affiliation{%
  \department{Institute for Visualisation and Interactive Systems}
  \institution{University of Stuttgart}}
\email{andreas.bulling@vis.uni-stuttgart.de}

% \email{e-mail address}

% \author{ANONYMISED}

% % The default list of authors is too long for headers.
% \renewcommand{\shortauthors}{}

\begin{abstract}
\input{01_abstract}

\end{abstract}



%
% The code below should be generated by the tool at
% http://dl.acm.org/ccs.cfm
% Please copy and paste the code instead of the example below.
%
% \begin{CCSXML}
% <ccs2012>
%  <concept>
%   <concept_id>10010520.10010553.10010562</concept_id>
%   <concept_desc>Computer systems organization~Embedded systems</concept_desc>
%   <concept_significance>500</concept_significance>
%  </concept>
%  <concept>
%   <concept_id>10010520.10010575.10010755</concept_id>
%   <concept_desc>Computer systems organization~Redundancy</concept_desc>
%   <concept_significance>300</concept_significance>
%  </concept>
%  <concept>
%   <concept_id>10010520.10010553.10010554</concept_id>
%   <concept_desc>Computer systems organization~Robotics</concept_desc>
%   <concept_significance>100</concept_significance>
%  </concept>
%  <concept>
%   <concept_id>10003033.10003083.10003095</concept_id>
%   <concept_desc>Networks~Network reliability</concept_desc>
%   <concept_significance>100</concept_significance>
%  </concept>
% </ccs2012>
% \end{CCSXML}

% \ccsdesc[500]{Computer systems organization~Embedded systems}
% \ccsdesc[300]{Computer systems organization~Redundancy}
% \ccsdesc{Computer systems organization~Robotics}
% \ccsdesc[100]{Networks~Network reliability}


% \begin{teaserfigure}
%   \includegraphics[width=\textwidth]{sampleteaser}
%   \caption{This is a teaser}
%   \label{fig:teaser}
% \end{teaserfigure}

\maketitle

\input{02_intro}

\input{03_relatedwork}

\input{05_method}

\input{04_dataset}

\input{06_evaluation}

\input{07_discussion}
\input{08_conclusion}

\bibliographystyle{ACM-Reference-Format}
\bibliography{references}

\end{document}

%% file: 01_abstract.tex
Recent methods to automatically calibrate stationary eye trackers were shown to effectively reduce inherent calibration distortion.
However, these methods require additional information, such as mouse clicks or on-screen content.
We propose the first method that only requires users' eye movements to reduce calibration distortion in the background while users naturally look at an interface.
Our method exploits that calibration distortion makes straight saccade trajectories appear curved between the saccadic start and end points.
We show that this curving effect is systematic and the result of distorted gaze projection plane.
To mitigate calibration distortion, our method undistorts this plane by straightening saccade trajectories using image warping.
We show that this approach improves over the common six-point calibration and is promising for reducing distortion.
As such, it provides a non-intrusive solution to alleviating accuracy decrease of eye tracker during long-term use.

 % only using eye movement analysis.

% Slides: \url{https://docs.google.com/presentation/d/1P-bGR8mzPpIMlihsdYy-OE0P3drLBwXbtHXKHgG7Bwk/edit?usp=sharing}

% Eye tracking allows for numerous advance HCI applicaions, however, calibrating an eye tracker can be combersome. 
% Recent efforts on continuous and implicit calibration of eye tracker offer an promising solution to gaze error accumulated after a long-term use. 
% However, such techniques oftentimes rely on mouse, keyboard events or screen saliency, which can be low-efficient or error-prone. 
% This paper presents the first self-calibration technique based on saccadic movements.
% It provides an innovative prospective to overcome the previous challenges. 
% By considering results of an uncalibrated eye tracker form a distorted gaze estimation surface, we cast the calibration into a rectification problem of the distorted gaze estimation surface. 
% We perform a user study to understand human saccadic movement behaviors [with different lengths and angles].
% We then leverage the [fact] that human saccades generally follow a straight line and exploit the deviated gaze points to correct the gaze estimation surface.
% Our evaluation demonstrates this method is effective, low-cost, and able to continuously improve the gaze estimation without either interaction or saliency information. 
% This method opens up a new avenue for eye tracker self-calibration, and it paves the way for practical gaze-aware HCI applications.

%% file: 02_intro.tex
\section{Introduction}

Eye tracking is flourishing given recent advances in hardware and software~\cite{huang2017screenglint,zhang2018training} as well as given increasing demands for mainstream applications, such as gaming or foveated rendering.
%It also becomes increasingly pervasive on personal devices, such as laptops, tablets and even phones. 
To achieve high tracking accuracy, eye trackers need to be calibrated to each individual user prior to first use.
During calibration, a gaze projection plane is estimated by asking the user to fixate at predefined locations on a computer screen \cite{duchowski2017eye} or to follow a moving dot~\cite{pfeuffer2013pursuit}.
While high eye tracking accuracy is achieved right after calibration, significant accuracy decrease was demonstrated during use~\cite{sugano2015self}, due to changes in users' head pose, relative position between screen and eye tracker, and other factors~\cite{blignaut2016idiosyncratic}.
We refer to the mapping from estiamted gaze point onto ground truth as \textit{calibration distortion}.
% Calibration drift describes the accumulating deterioration in eye tracking accuracy after an initial manual calibration was performed.
To address this problem, previous works proposed 
% to address calibration drift 
post-hoc correction
% , e.g.\ through additional
 correction~\cite{vspakov2014real},
% filtering \cite{vspakov2012comparison} 
or to embrace it in the design of error-aware gaze interfaces~\cite{barz18_etra}.
%\michael{I am not quite sure whether filtering can be a solution for ill-calibration, because filtering, I guess, is only to improve precision (maybe caused by visual noise/jittering), not to improve accuracy. Maybe talk about post-hoc correction instead.}
%\andreas{sentence on what the problem of these methods is.}
However, these approaches only alleviate the symptoms and do not address the problem directly.
%However, all of these methods are based on the on-screen location of ground truth gaze points, \andreas{true? also probably too fuzzy. which requires a secondary input.}

Another line of work introduced the idea of \textit{self-calibration}, i.e.\ continuous recalibration in the background while the eye tracker is being used~\cite{sugano2015self}.
While this approach was shown to be effective 
% in reducing calibration drift
, current self-calibration methods either assume correlated, secondary user input, such as mouse clicks~\cite{huang2016building} or touch input~\cite{zhang2018training}, or require information about on-screen content to compute saliency maps~\cite{sugano2015self}.
%However, existing methods either require secondary input information to provide the gaze location through information, such as a mouse click and a keypress, and the visual saliency of the user's egocentric view.
While user input and gaze are often correlated \cite{huang2016building,sugano2015appearance}, this correlation is far from perfect and cannot be guaranteed.
In addition, collecting sufficient and high-quality interaction data remains challenging.
Similarly, while saliency maps can predict likely on-screen gaze locations~\cite{sugano2015self}, the reliability of these predictions can be low and the computational cost of computing them can cause problems for real-time operation.

\begin{figure}[t!]
    \centering
    \includegraphics[width=.95\columnwidth]{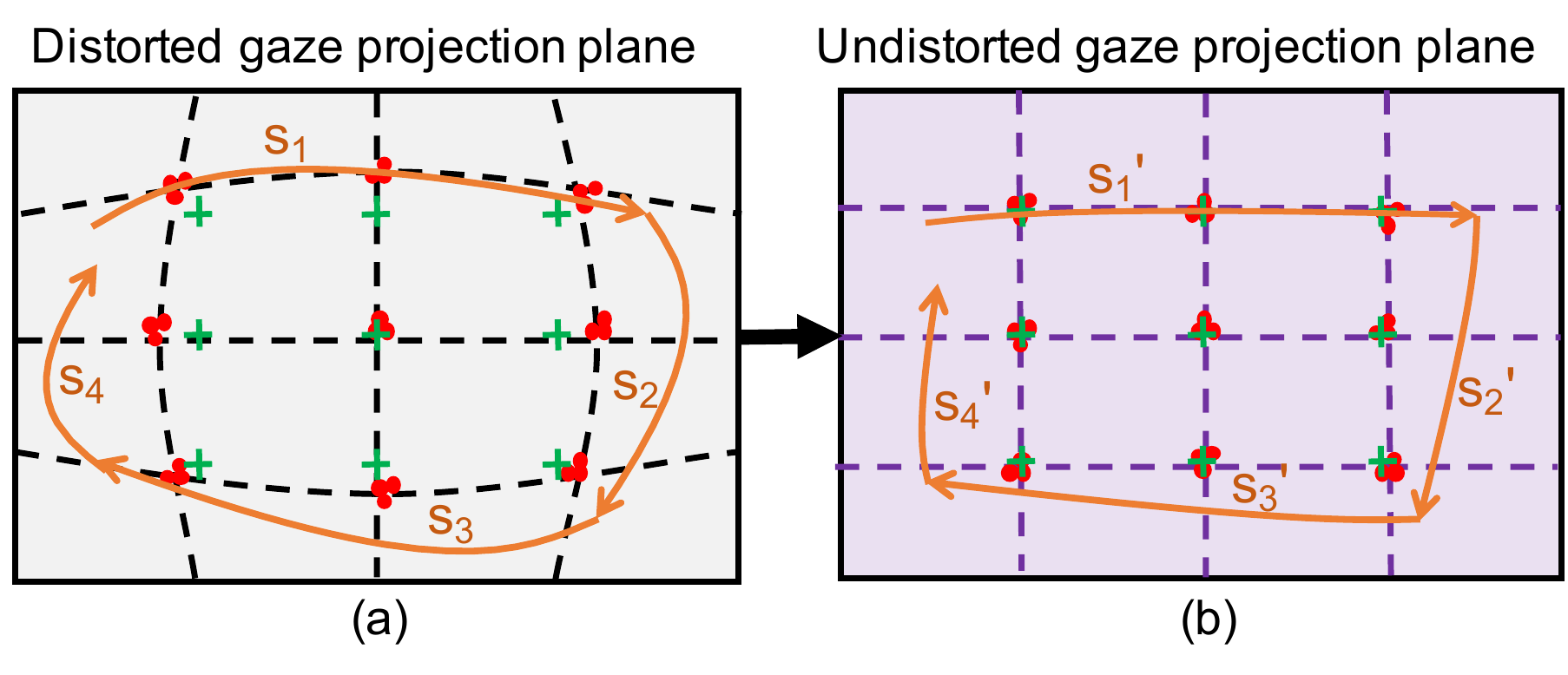}
    \caption{We correct the distorted gaze projection plane (black dash lines) to the undistorted gaze projection plane (purple dash lines) by minimizing the curvature of curved saccade trajectories (orange). By doing so, gaze points (red) are transformed closer to ground truth locations (green). 
    %\andreas{terms: warped vs undistort vs rectify}
    }
    \label{fig:teaser}
\end{figure}

To address these limitations we propose \textit{SacCalib} -- the first calibration distortion reduction method that only requires information about a user's eye movements.
That is, without the need for secondary user input or time-consuming computation of saliency maps for on-screen content.
%In this paper, we present the first eye-only self-calibration method.
%It gets rid of the constraint from the secondary inputs, thus provides a low-cost alternative to the self-calibration of eye tracker.
Saccade trajectories recorded by a calibrated eye tracker are nearly straight between the saccadic start and end points, i.e.\ form a regular
%\andreas{not sure if the right work, opposite of distorted: flat}
%\michael{how about: regular/even/uniform? or identity mapping?; the deformed one is also 2D -- still flat}
gaze projection plane (see Figure~\ref{fig:teaser}b).
%\andreas{not sure about this gaze estimation surface}
%\michael{how about "gaze estimation mapping" that transforms the gaze estimates to the ground truth gaze locations.}
%\andreas{``gaze mapping function'' is used in the abstract. Terminology is secondary, just make sure people understand what you mean.}
The key observation that our method builds on is that
% an uncalibrated eye tracker 
% the drift from ideal calibration
\textit{without calibration straight saccade trajectories appear curved}, resulting in a \textit{distorted gaze projection plane} (see Figure~\ref{fig:teaser}a).
% \andreas{add a and b to teaser}
%As shown in this work, this curving effect is systematic and specific to a particular head pose, eye tracker, and screen geometry.
% As shown in this work, this curving effect is systematic.
We see that this curving effect is systematic.
% and the result of an uncalibrated and thus distorted gaze mapping function.
% directly linked to the error-prone gaze mapping function.
%In a nutshell, our method leverages the straight saccadic eye movements to correct the \textit{gaze estimation surface} (i.e. the projected surface of the gaze estimates) and therefore reduces the eye tracking error. 
%As shown in Figure~\ref{fig:teaser}, saccades on the original gaze estimation surface from an ill-calibrated eye tracker appear curved. 
That is, by observing multiple saccades performed between different on-screen locations, and by jointly minimizing saccade curvatures and thus undistorting the corresponding gaze projection plane, calibration distortion can be reduced.
%, the resulting warped surface projects the gaze estimates closer to the ground truth gaze locations.
%In other words, we cast the problem of eye tracker calibration into a plane undistortion problem assuming straight saccade trajectories. 

% A key challenge for our method is that saccadic eye movements can be ballistic and, while straight in principle, may have a base curvature. 
A key challenge for our method is that saccadic eye movements, while straight in principle, can be curved. 
However, as suggested in~\cite{godijn2004relationship,moehler2014effects,moehler2015influence}, natural saccade curvature can be reduced given sufficient preparation time for performing a saccade.
%, i.e.\ when a user has sufficient preparation preceding a saccade, or when the user shifts his/her covert attention to the saccadic target in advance.
%As a first step towards eye-only self-calibration, the current study focuses on only the saccades with a high chance to be straight.
% \andreas{one sentence on the implication of this for our method; how do we deal with this challenge?}
We thus design an experiment paradigm to capture saccades under such condition and demonstrate the effectiveness of eye-only calibration distortion reduction.
% We also study the impact on our calibration technique from the relevant saccadic attributes, including the saccade direction and amplitude. 
%\andreas{not only study the impact -- you identify the impact of saccae direction and amplitude on calibration quality and identify the best parameters etc}
As such, our method opens up an exciting new avenue for eye tracker self-calibration and also paves the way for numerous practical gaze-aware HCI applications that do not require frequent, cumbersome, and time-consuming explicit eye tracker recalibration.
%\michael{I think "do not require explicit eye tracker recalibration" would be more appropriate; we are not going to show it can calibrate the eye tracker from scratch.}

The contributions of our work are two-fold.
%First, we study the impact of saccade direction and amplitude on saccade curvature.
First, we propose the first eye-only calibration distortion reduction technique based only on saccadic eye movements.
In contrast to current self-calibration methods, our method neither requires secondary user input, such as mouse clicks, nor expensive processing of on-screen content.
%\andreas{same sentence as in the abstract about benefits of this approach.}
%Second, we investigate the saccade attributes and their impacts on the performance of the eye-only calibration.
Second, we evaluate our method on a newly collected, 10-participant dataset of around 3,000 on-screen saccades, which we will release to the research community upon acceptance.
Through this evaluation, we provide insight into the key issues of eye-only calibration.

%% file: 03_relatedwork.tex
% group 1 and 2

\section{Related Work}

Our work is informed by research on natural curvature of saccadic eye movements and prior work on eye tracker self-calibration.

\subsection{Curvature of Saccadic Eye Movements}

The reason for saccade curvature~\cite{viviani1977curvature,van2010recent} is still an open research question~\cite{kruijne2014model,smit1990analysis,van2006eye}.
% Existing literature suggested causes for saccade trajectory curvatures~\cite{viviani1977curvature,van2010recent}, including 
Potential causes include oculomotor inhibition~\cite{doyle2001curved,tipper1997selective}, saccadic latency~\cite{mcsorley2006time}, top-down selection processes~\cite{van2006eye}, and residual motor activity~\cite{rizzolatti1987reorienting,wang2011aftereffects}.
Oculomotor inhibition denotes competing saccade programs for the target and task-irrelevant distractor, which cause saccade deviation away from or toward the distractor~\cite{mcpeek2003competition}.
% McPeek et al. shown that saccadic eye movement can either curve away from or toward the distractor~\cite{mcpeek2003competition}. 
It may also be related to saccade attributes, such as saccade (early/late) stages~\cite{mcsorley2006time,walker2006control} or \textit{saccadic latency} between saccade starting indicator and saccade onset~\cite{ludwig2003target,mcsorley2006time,moehler2015influence}.
On the other hand, it was argued that saccade curvature may stem from top-down selection processes of the target~\cite{van2006eye}. 
An unresolved competition among visual targets results and a clear goal-directed orienting may lead to different saccade curvatures.
In addition, the second saccade in consecutive saccades could curve away from the initial fixation~\cite{megardon2017trajectory}.
This is regarded as residual motor activity~\cite{wang2011aftereffects}.
% On the other hand, Van der Stigchel et al.\ argued that saccade curvature may stem from \textit{top-down selection processes} of the target~\cite{van2006eye}. 
% That is, unresolved competition in the oculomotor system among visual targets results in the deviations towards the distractor, while the clear goal-directed orienting leads to the deviations away from the distractor.
% Megardon et al. further studied the saccade sequences and found that the second saccades could curve away from the initial fixation~\cite{megardon2017trajectory}.
% This leads to another cause of saccade curvature: residual motor activity of saccades, which forces successive saccades to curve along the direction of the previous saccade~\cite{wang2011aftereffects}.

It is worth noting that saccade curvature has been observed to decline or even vanish with an increase of \textit{movement preparation time} preceding the saccades~\cite{godijn2004relationship,moehler2014effects,moehler2015influence}.
This implies that sufficient time 
% or covert attention shifts prior to the next saccade target 
allows for the completion of top-down selection process among targets and thus reduces oculomotor inhibition and leads to straight saccades.
% \andreas{last sentence: not clear how this links to the sentence before or the current work in general. covert attention not mentioned before.}

\begin{figure*}
    \centering
    \includegraphics[width=1.\textwidth]{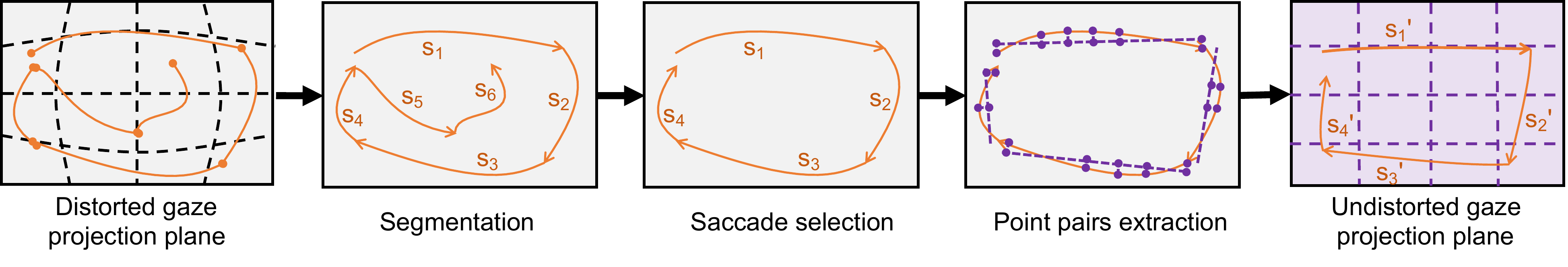}
    \caption{Our method segments the saccades, selects those with suitable properties (see main text for details), and extracts point pairs on the curved saccade trajectories and its straight counterpart. It then corrects the distorted gaze projection plane to the undistorted projection plane by jointly minimizing saccade curvatures, thereby effectively reducing calibration distortion.}
    \label{fig:overview}
\end{figure*}

\subsection{Eye Tracker Self-Calibration}

% \andreas{in the abstract and intro we refer to two prior lines of work and their drawbacks (secondary input and additional camera). Make sure this is picked up here again and/or modify the corresponding locations in abstract/intro. ALso make sure terminology is consistent: ``interaction cues'' and ``secondary input''}

Most commerial eye trackers relies on explicit calibration, which requires users to fixate on on-screen calibration points.
For example, Tobii eye tracker applies a six-point calibration procedure that shows four points on screen corners and two in center.
Although this procedure is usually short, a frequent re-calibration to maintain eye tracking accuracy is intrustive to users. 
Therefore, increasing research demands has rised for eye tracker implicit self-calibration.

Eye tracker self-calibration can be categorized into post-hoc correction~\cite{vspakov2014real,barz18_etra}, saliency-based~\cite{chen2015probabilistic,sugano2015self,sugano2013appearance,wang2016deep} and eye movement based approaches~\cite{huang2016building,khamis2016textpursuits,papoutsaki2016webgazer,pfeuffer2013pursuit,sugano2015appearance}. 
The first one conducts post-hoc correction of gaze estimation without modifying the gaze projection plane, while the other two aim to overcome calibration distortion by correcting the plane distortion as this study.
Specifically, saliency-based method extracts \textit{saliency map} of either screen image or user's egocentric view and then maps eye features into
% \andreas{unclear: build  associations between gaze and objects} in the 
image coordinate by assuming that the user is likely to look at the most salient locations.

Early works used bottom-up saliency maps that model the influence of low-level image attributes, such as edge, shape, and color~\cite{itti1998model,koch1987shifts}, as well as the high-level image semantics~\cite{huang2015salicon}, including objects~\cite{xu2014predicting}, human faces~\cite{sugano2015self}, gaze location of a person~\cite{gorji2017attentional} or multiple persons~\cite{fan2018inferring}.
More recent works investigated top-down saliency maps that account for goal-oriented or task-controlled visual attention and cognitive processes~\cite{borji2012probabilistic,huang2018predicting,peters2007beyond}. 
However, calculating saliency maps for each camera frame is computational expensive and saliency maps can be highly inconsistent with users' actual visual attention~\cite{judd2009learning}.

In contrast, eye movement based approaches exploit secondary user input.
% interaction cues \andreas{what do you mean with this? interaction cues, consistency? that are consistent with eye gaze}.
Specifically, conventional approaches include fixation-based~\cite{huang2016building,papoutsaki2016webgazer,sugano2015appearance} and pursuit-based~\cite{khamis2016textpursuits,pfeuffer2013pursuit}. 
The general assumption of fixation-based methods is that the user looks at the interaction location, such as locations of mouse-clicks~\cite{huang2016building,papoutsaki2016webgazer,sugano2015appearance}, mouse movements~\cite{huang2012user,papoutsaki2016webgazer}, and key presses~\cite{huang2016building}.
In contrast, pursuit-based calibration relies on the movement correlation between visual stimuli and eye gaze, and the moving stimuli can be a specific cursor~\cite{pfeuffer2013pursuit}, natural texts~\cite{khamis2016textpursuits}, or an object in games~\cite{tripathi2017statistical}.
Despite the close link between eye movement and user input, eye movement-based calibration is limited by input sparsity in real use, and pursuit-based calibration requires dynamic interfaces. 
Therefore, eye-only calibration that does not require visual scene or user input can be highly beneficial to addressing the limitations of conventional calibration techniques. 

% \michael{compare with mouse movement based calibration.}

%% file: 05_method.tex
\section{Reducing Distortion via Warping}
% \section{Reducing Calibration Drift via Image Warping}

% \andreas{general comment: you use ``the'' rather excessively in front of nouns (something Germans are usually know for)}

To correct the distorted gaze projection plane, we first identify and segment saccades using the velocity threshold method (I-VT)~\cite{salvucci2000identifying} following common practice~\cite{arabadzhiyska2017saccade}. 
We then select suitable saccades and extract all pairs of gaze points along the curved and straight saccade trajectories.
% as well as the straight saccadic trajectories.
These point pairs are input to an image warping technique that uses moving least squares~\cite{schaefer2006image} to correct curved saccade trajectories to straight.
% the \textit{control points} (i.e.\ gaze points along the original saccade trajectory) to the target locations on the corrected saccade trajectory.
% \andreas{last sentence not clear: original saccade clearly defined? control points clear? transform to corrected trajectory seems illogical as the corrected trajectory is the goal.}
% \andreas{I'd always speak about saccade trajectory, that is more precise}
As a result, the distorted gaze projection plane is undistorted.
 % and thus calibration drift is reduced.
\autoref{fig:overview} shows the method overview.

\subsection{Extracting Point Pairs for Warping}

\begin{figure}
    \centering
    \includegraphics[width=0.9\columnwidth]{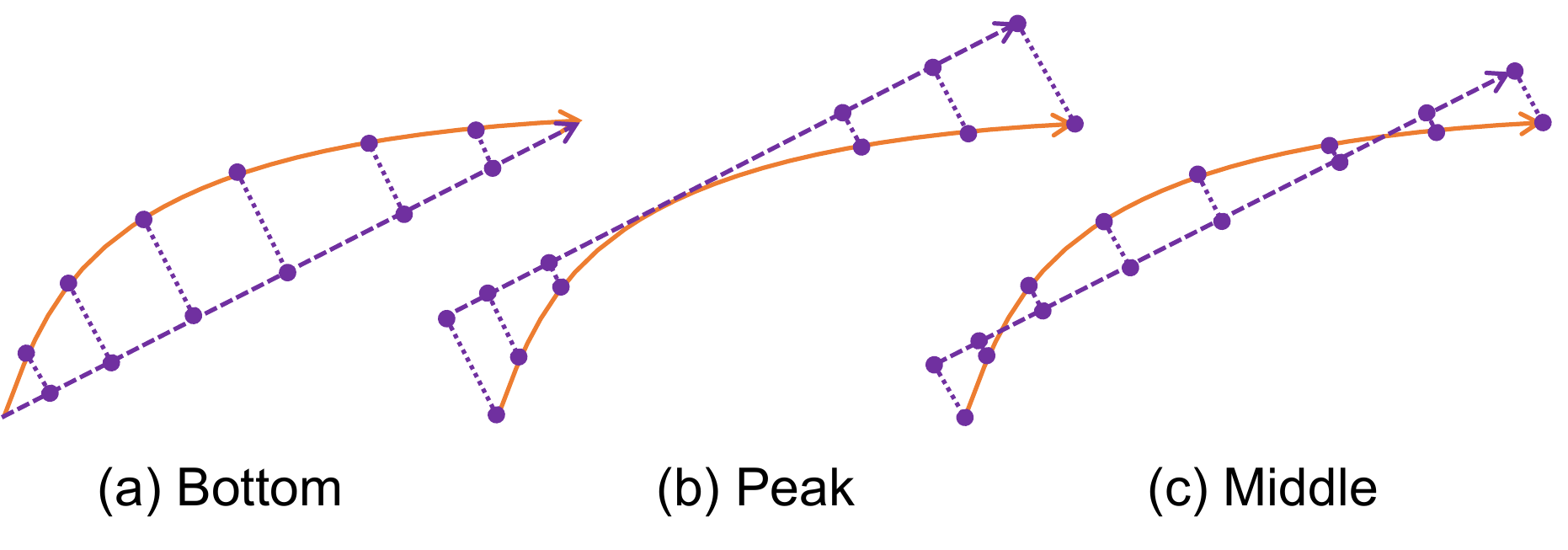}
    \caption{%Illustration of saccade correction that removes the curvature due to distortion. We can correct
    Curved saccade trajectory (orange curve) is corrected to straight saccade (purple dash line) by (a) projecting to the line that connects two endpoints; or (b) with an additional shift to the peak of the curved saccade; or (c) to the middle toward the peak. Purple dots indicate gaze points.}
    \label{fig:correction}
\end{figure}

% \andreas{we have two terms now: curved and distorted. should curved be used for trajectories, maybe, and distorted for the projection plane?}
The basis of our eye-only calibration method lies in the rectification of the curved saccade trajectories to (assumed) straight trajectories.
However, the solution to correcting a curved trajectory is not unique. 
\autoref{fig:correction} shows three potential projections with different offsets from the saccade endpoints.
By maintaining the saccade direction pointing from the saccade start to the end, we can a) project the curved saccade trajectory to the straight line that connects the endpoints of the saccade; we can also b) shift it to the peak of the distorted saccade, so that they just touch each other, or c) shift it to the middle toward the peak.
We adopt the projection with shift to the middle in our method.
Let $\langle{p_{1},...,p_{m}}\rangle$ and $\langle{p'_{1},...,p'_{m}}\rangle$ be the gaze points on all the distorted saccades, and their projection counterparts on the corrected saccades, respectively. 
The point pairs that control the warping of the gaze project plane can be represented by $\langle{\{p_{1},p'_{1}\},...,\{p_{m},p'_{m}\}}\rangle$.

\subsection{Undistorting the Gaze Projection Plane}

% The core of the proposed method is to warp the distorted gaze estimation surface to improve eye tracking accuracy.
Inspired by a popular image warping technique used in computer vision~\cite{schaefer2006image}, we 
% apply the gaze points from the distorted saccades as control points and 
undistort the gaze projection plane by minimizing the distance between the pairs of gaze points on curved and straight saccade trajectories. 

There are three desired properties of undistorting the gaze projection plane. 
First, it should reduce the gap between gaze point pairs.
Second, it should produce smooth deformations, i.e.\ the area among different gaze point pairs should be smooth.
Third, it should preserve the original relative geometry, as e.g.\ a point on the left side of a saccade is expected to stay on the left after warping. 

To this end, given a point $v$, we solve for the affine transformation $l_v(x)$ that minimizes 
\begin{equation}
\sum _{i}^{m}{w_{i}|l_{v}(p_{i})-p'_{i}|^2}
\label{equ:obj}
\end{equation}

\noindent
where $p_{i}$ and $p'_{i}$ are the point pair that controls warping and $w_{i}$ are the weights that control the impact of each point pair on the transformation of point $v$. 
Intuitively, the weights should be inversely related to the distance from the input points to achieve the smoothness of transformation, thus we define it as $w_{i}=1/|p_{i}-v|$.
% \begin{equation}
% w_{i}=\frac{1}{|p_{i}-v|^{\note{2\alpha}}}
% \end{equation}

As pointed out in \cite{schaefer2006image}, the affine transformation $l_v(x)$ should consist of a linear transformation and a translation, but the translation component can be substituted by referring to the weighted centroids of the point pairs.
That is, \autoref{equ:obj} can be rewritten in term of the linear matrix $M$.
\begin{equation}
\sum _{i}^{m}{w_{i}|\hat{p}_{i}M-\hat{p}'_{i}|^2}
\end{equation}

\noindent
where $\hat{p}_{i}=p_{i}-\sum _{i}w_{i}p_{i}/\sum _{i}w_{i}$ and $\hat{p}'_{i}=p'_{i}-\sum _{i}w_{i}p'_{i}/\sum _{i}w_{i}$ are the normalized point pair by their weighted centroids, respectively.
Depending on the form of matrix $M$, we can fine-control the transformation characteristics. 
Specifically, using the general form of matrix $M$ results in a fully affine transformation, which contains non-uniform scaling and shear.
Restricting matrix $M$ to a similarity transformation that only includes translation, rotation, and uniform scaling better preserves angles on the original plane.
Further restricting matrix $M$ to a rigid transformation that excludes scaling can maintain the relative geometry after warping.
We therefore use a rigid transformation form of matrix $M$.
Please refer to \cite{schaefer2006image} for more information.

To speed-up the computation, we approximate the full plane with a fine-grained grid with a quad size of 25 pixels and only apply the deformation to each vertex in the grid.
We then perform bilinear interpolation to fill each pixel in the quads according to the values of the four corresponding  corners. 
The computation time is thus linear in the number of vertices in the grid.

Due to the potentially contradictory warping controlled by 
% \andreas{English: mapping assigned} 
different point pairs, the undistorted projection plane may suffer from undesirable fold-back, where a point on one side of a line may be mapped to the other side.
Therefore, we apply a post-hoc spatial smoothing on the resulting transformation, by using a normalized box filter with a blurring kernel size of 5 on the grid data.

% $M=argmin_{M} \sum_{j}\sum_{L_i}{Proj(M(x_i),L_j)}$
% how to perform a optimization for this?

\subsection{Selecting Saccades}

Although the above method can improve gaze accuracy by correcting the distorted saccades, in practice not all saccades are purely deformed by the distorted gaze projection plane.
Instead, saccadic eye movements can be noisy, sometimes too short, and mixed with jittering~\cite{farmer1991optimal}. 
% \andreas{add refs}
The ideal saccade candidates for our method are those that are straight by nature. 
In addition, they should be long enough to impose an valid effect on the plane correction.
Therefore, we filter saccades before applying only the suituable candidates to undistort the gaze projection plane.
Specifically, we construct a random forest classifier that selects saccades based on multiple attributes.
We start with the discussion about the curvature measure, given its close link to our core idea.

\subsubsection{Measure of saccade curvature}

\begin{figure}[t]
    \centering
    \includegraphics[width=0.9\columnwidth]{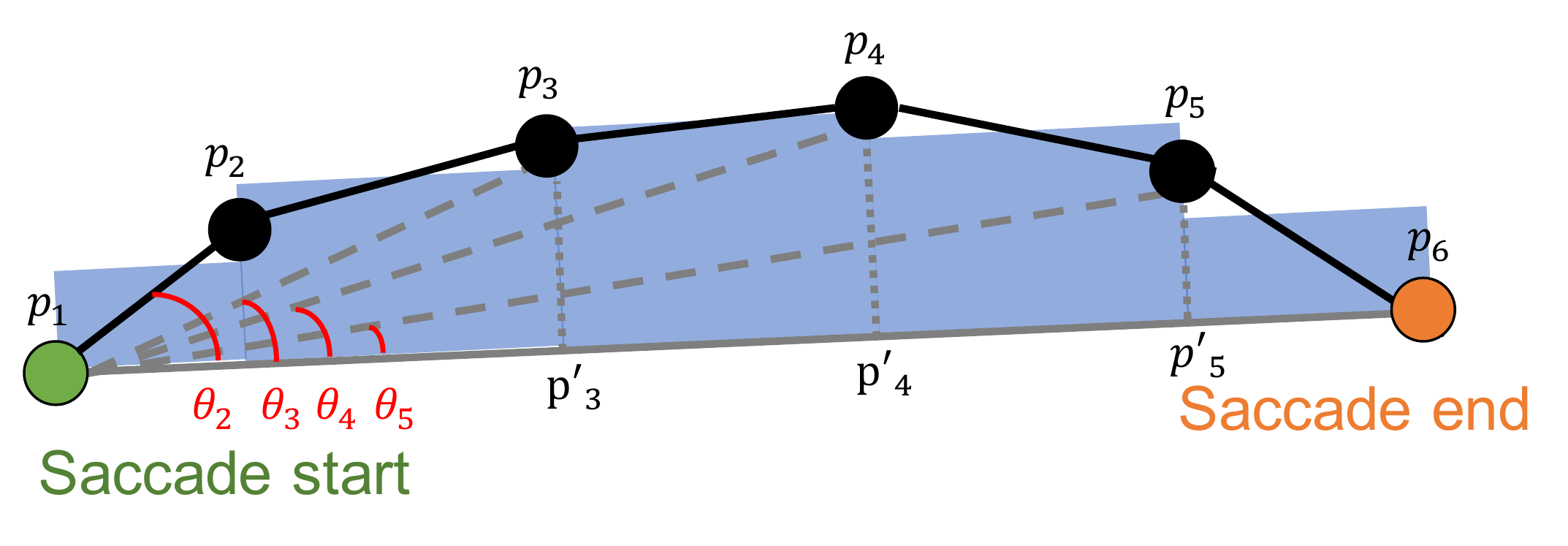}
    \caption{Illustration of the area- and direction-based measures for saccade curvature. The former one computed the area covered by the curved saccade trajectory, while the latter is the average angle of gaze points w.r.t.\ the endpoints.}
    \label{fig:metrics}
\end{figure}

Measures of saccade curvature can be direction-based, distance-based, area-based, or curve fitting-based~\cite{tudge2017setting,van2006eye}.
These measures can be computed with respect to the location of the target or the endpoint of saccade.
As we aim for eye-only calibration, where the actual target is agnostic to the method, the current study focuses on the endpoint-based measures.
% In this study, 
We compute the area-based curvature as well as the direction-based curvature, as suggested in~\cite{tudge2017setting}.
Specifically, the \textit{area-based curvature} measures the area between the saccade trajectory and the straight line from saccade start to saccade end.
As shown in~\autoref{fig:metrics}, let $p_{i}$ in $\langle{p_{1},...,p_{m}}\rangle$ denote the $i$-th point on a saccade with $m$ points, the area-based curvature, $CurvatureArea$, is computed by

\begin{equation}
CurvatureArea=\sum _{i=2}^{m}\frac{|p_{i}p'_{i}|+|p_{i-1}p'_{i-1}|}{2}|p'_{i}p'_{i-1}|
\label{equ:CurvatureArea}
\end{equation}

The \textit{direction-based curvature} denotes the average angles formed by lines from saccade start to each gaze point on the saccade, with respect to the straight line connecting two saccade endpoints. 
Let $\theta_{i}$ be the $i$-th angle, i.e. $\angle{p_{i}p_{1}p_{m}}$, the direction-based curvature, $CurvatureAngle$, is quantified by

\begin{equation}
CurvatureAngle=\frac{1}{m}\sum _{i=2}^{m-1}\theta_{i}
\label{equ:CurvatureAngle}
\end{equation}

\subsubsection{Identifying suitable
% \andreas{was useful becore; I suggest you replace everywhere with: suitable} 
saccades using a data-driven approach}

Apart from saccade curvature, we also consider other types of attributes, including amplitude, orientation, velocity, and timing (see \autoref{tab:features}). 
The reason that we include these attributes stems from findings about natural saccade curvature in human vision research. 
For instance, vertical and oblique saccades were found to be more curved than the horizontal ones~\cite{bahill1975neurological,kowler2011eye,yarbus1967eye}.
Similarly, saccade amplitude can be pertinent~\cite{van1987skewness}. 
Besides, saccadic latency was also found to be related to the saccade curvature~\cite{ludwig2003target,mcsorley2006time,moehler2015influence}.
As such, the saccade orientation, amplitude, and timing can be good indicators for its natural curvature.
In addition, given the close link between saccade curvature and its velocity profile, we also include the velocity attributes.

We extract and use these attributes as input to a random forest classifier to predict the suitability for correction of each saccade. 
%We use the default parameters of the classifier.
We set the number of attributes in each tree to 5, the number of trees in the forest to 20, the maximum depth of the tree to 50 to reduce the risk of overfitting due to our medium-size training data.
% Additionally, to reduce the risk of overfitting due to our medium-size training data, we constrain the maximum depth of the tree to 50.

\renewcommand*{\arraystretch}{1.2}
\begin{table}[t!]
\small
\begin{tabular}{p{1.35cm}|p{6.45cm}}
\toprule
\textbf{Type} & \textbf{Measure of the attributes}\\
\midrule
Curvature   (2) & $CurvatureArea$ is an area-based measure and $CurvatureAngle$ is a direction-based measure according to \autoref{equ:CurvatureArea} and \autoref{equ:CurvatureAngle}, respectively. \\
\midrule
Amplitude   (2) & $Amplitude$ measures the direct distance between saccade start and end, while $Length$ measures the sum distance between all consecutive gaze points on a saccade.\\
\midrule
Orientation (2) & $Direction$ denotes the saccade angle with respect to the horizontal line; $Turn$ presents the number of large direction change (>90$^{\circ}$).\\
\midrule
Velocity    (2) & $Velocity$ measures the overall velocity of all the points on a saccade, and $VelocityMax$ delineates the peak velocity on it.\\
\midrule
Timing      (2) & $Latency$ measures the interval between the vanishing of the last target and the saccade start, and $Duration$ measure the saccade duration.
\\
\bottomrule
\end{tabular}
\vspace{0.2cm}
\caption{We extract 10 saccade attributes.
% \andreas{terminology: attributes vs features} 
% to describe curvature, amplitude, orientation, velocity, and timing. 
% The number in parenthesis shows the number of attributes in the corresponding type of attribute.
The number of attributes in each type is shown in the parenthesis.}
% \caption{We extracted a total of 52 eye movement features to describe a user's eye movement behaviour. The number of features per category is given in parentheses.}
    \label{tab:features}
\end{table}

\subsubsection{Acquiring training data}

To obtain the training samples 
% for the random forest classifier
, we perform a leave-one-saccade-out procedure on the session-based recording data.
That is, we start with using all the saccades for plane correction. 
We then iteratively leave out one random saccade if the removal improves the overall accuracy. 
We stop the procedure until no more improvement can be achieved. 
This gives us a \textit{suitable saccade subset} that contributes to the plane correction. 

In our setting, the remaining saccades from different sessions take approximately 10\% to 30\% of the original saccades.
To create a balanced training set for the random forest classifier, we perform a random downsampling on the major class (i.e. \textit{unsuitable saccade subset}). 
Finally, we group the suitable and unsuitable subsets of saccades from different sessions and participants for training.
% to form the final saccade set for training.

\subsection{Segmenting Saccadic Eye Movements}

Eye tracking data mainly contains saccades, fixations, and blinks. 
We follow previous practice to segment saccades, according to the velocity profile of the eye movements~\cite{arabadzhiyska2017saccade}. 
% \andreas{ref?}

\subsubsection{Fixing the noisy eye tracking data}

Raw eye tracking can be noisy, due to eye blinks, poor tracking quality, motion blur, and infrared reflection on glasses. 
We first remove high-frequency jitter. That is, medium ($\sim$1$^{\circ}$) jerk-like eye movements at an abnormal frequency of around 100 Hz that occur frequently near the screen boundary.
We then use linear interpolation to fill missing data of short durations (<50 ms). 
Such data loss is likely a result from visual noise or tracking failure, for normal eye blink duration is around 100-140 ms~\cite{schiffman1990sensation}. 
We finally apply a low-pass filter~\cite{farmer1991optimal} to remove high-frequency noise.
% in the eye movement signals. 

\subsubsection{Segmenting saccades}

After the above preprocessing, we apply the I-VT method~\cite{salvucci2000identifying} to segment saccadic eye movements.
Specifically, we define three velocity thresholds as in~\cite{arabadzhiyska2017saccade,dorr2010variability}: a detection threshold $V_{d}$ (100$^{\circ}$/s), a starting velocity threshold $V_{a}$ (60$^{\circ}$/s), and a final velocity threshold $V_{f}$ (60$^{\circ}$/s).
The \textit{detection threshold} is used to identify the first gaze point whose velocity exceeds $V_{d}$ and we refer to it as the \textit{detection point}.
This threshold identifies a gaze point on a saccade with a safe margin. 
Since the actual saccade start is earlier than the detection point, we scan backward from the detection point and look for the saccade start, where  velocity begins to exceed $V_{a}$.
Similarly, we seek forward for the gaze point whose velocity begins to drop beyond $V_{f}$ and refer to it as the saccade end.
The values of these threshold parameters are in good agreement with prior studies~\cite{arabadzhiyska2017saccade,boghen1974velocity}.

%% file: 04_dataset.tex
\section{Collecting a Saccade Dataset}

% \michael{should experiment or method go first?}

% \andreas{In terms of logic this dataset is, first, to analyse saccadic curvature and the infuence of different saccade attributes on it (see intro and contribution statement). Second, it is also the basis, in combination with the obtained findings, for developing and evaluating the method.}

To study the feasibility and effectiveness of the proposed eye-only calibration method we collected a 10-participant dataset of saccades with different amplitudes and directions. 

% \textit{long} and \textit{short} saccades and 2) \textit{oblique} and \textit{axial} (i.e. horizontal/vertical) saccades. 

\subsection{Experiment Design}

\begin{figure}
    \centering
    \includegraphics[width=.8\columnwidth]{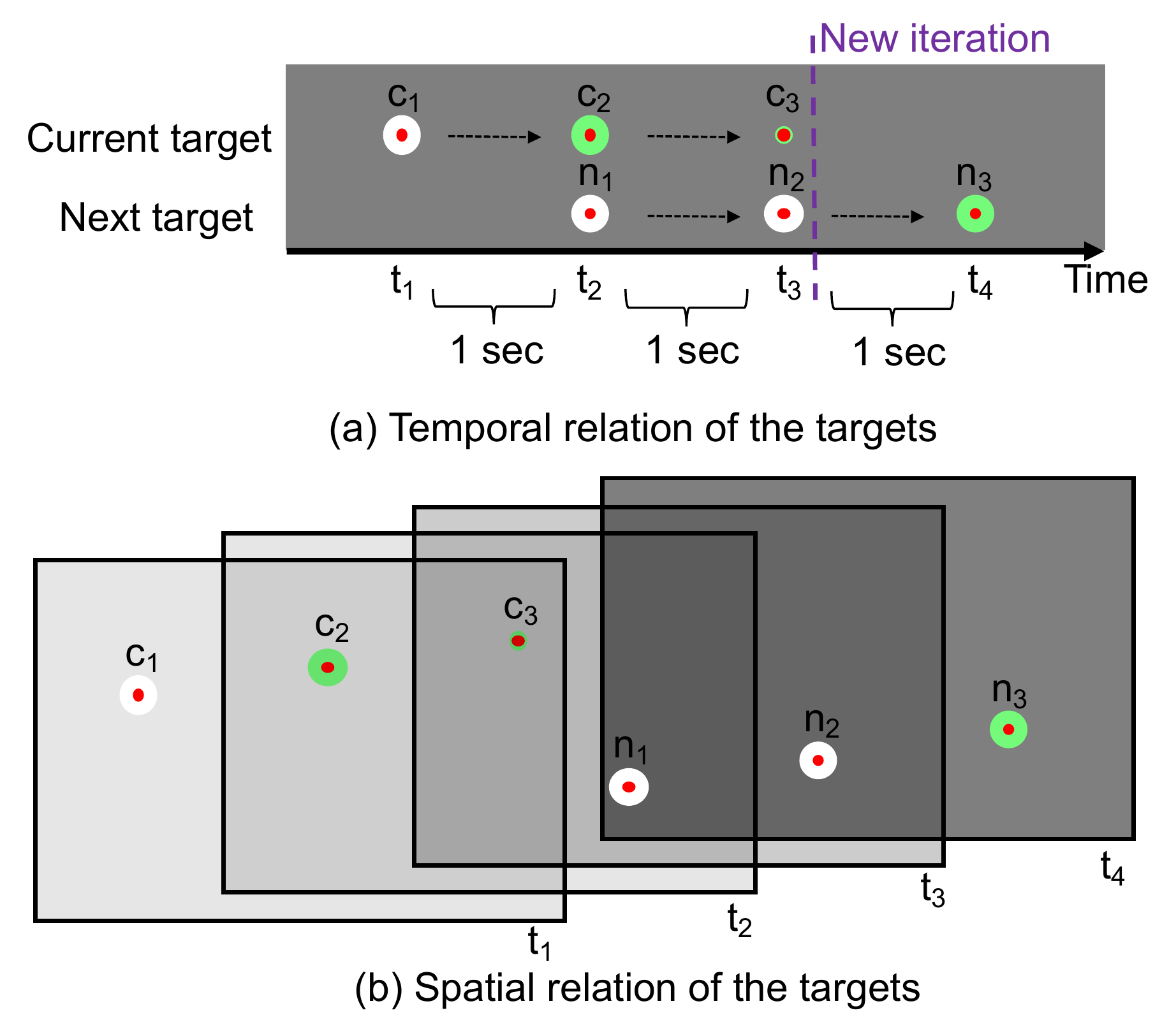}
    \caption{The (a) temporal and (b) spatial relation of the current and the next visual targets. The current target (in green) shrinks and disappears, but once it starts to shrinks the next target (in white) is shown in another random location defined by a five-by-five grid that evenly covers the screen.}
    \label{fig:design}
\end{figure}

Since the proposed method assumes straight saccades for the correction of the gaze projection plane,
%, we investigated different saccadic attributes that may bend the saccade trajectory. 
%Please be reminded that saccade curvature can be dynamic due to oculomotor inhibition~\cite{doyle2001curved,tipper1997selective}, top-down selection processes~\cite{van2006eye}, and residual motor activity~\cite{rizzolatti1987reorienting,wang2011aftereffects}.
we designed our data collection procedure 
%to minimize bending effects,
so as to capture natural straight saccades.
% and evaluate the core idea of the proposed method.
Specifically, we used a shrinking circle shown on a black screen to direct the saccadic eye movement of the participants. 
As shown in ~\autoref{fig:design}, the current visual target is shown initially in white with a red dot in its center, and it gradually turns green in one second and then shrinks and disappears in another second. 
Once the current (green) target starts to shrink, the next (white) target shows up.
When the current target shrinks to vanish, participants should make a saccade to the next target.
To minimize the impact of oculomotor inhibition and top-down selection processes, this study focuses on the scenario with no distraction for saccades.

Visual targets were shown in random locations on a five-by-five grid spread evenly across the screen, resulting in saccades with diverse directions and amplitudes. 
This was inspired by prior studies showing that horizontal saccades are more likely to be straight~\cite{yarbus1967eye}, while vertical and oblique saccades generally appear curved~\cite{kowler2011eye}. 
As suggested by Moehler and Fiehler~\cite{moehler2015influence}, the saccade curvature effect vanishes after one second saccade preparation time.
Therefore, one second before the current visual target disappeared, the next target already appeared in another location on the screen.
We encouraged participants to locate that next target using peripheral vision, i.e.\ to shift their covert attention toward it once it appeared. 
%Throughout the experiment, participants were instructed to fixate on the target circle.
%Eye blinks were only allowed while looking at the current target.
To help users discriminate between targets, the current target was always shown in green while the next one was shown in white.
% to help participants quickly discriminate between the targets and an indicator for the blink allowed period.

%The saccades we collected in this manner can produce the evaluation data to verify the previous findings given sufficient preparation time, and more importantly, facilitates us to study the eye-only calibration using oblique saccades.

\begin{figure}
    \centering
    \includegraphics[width=.8\columnwidth]{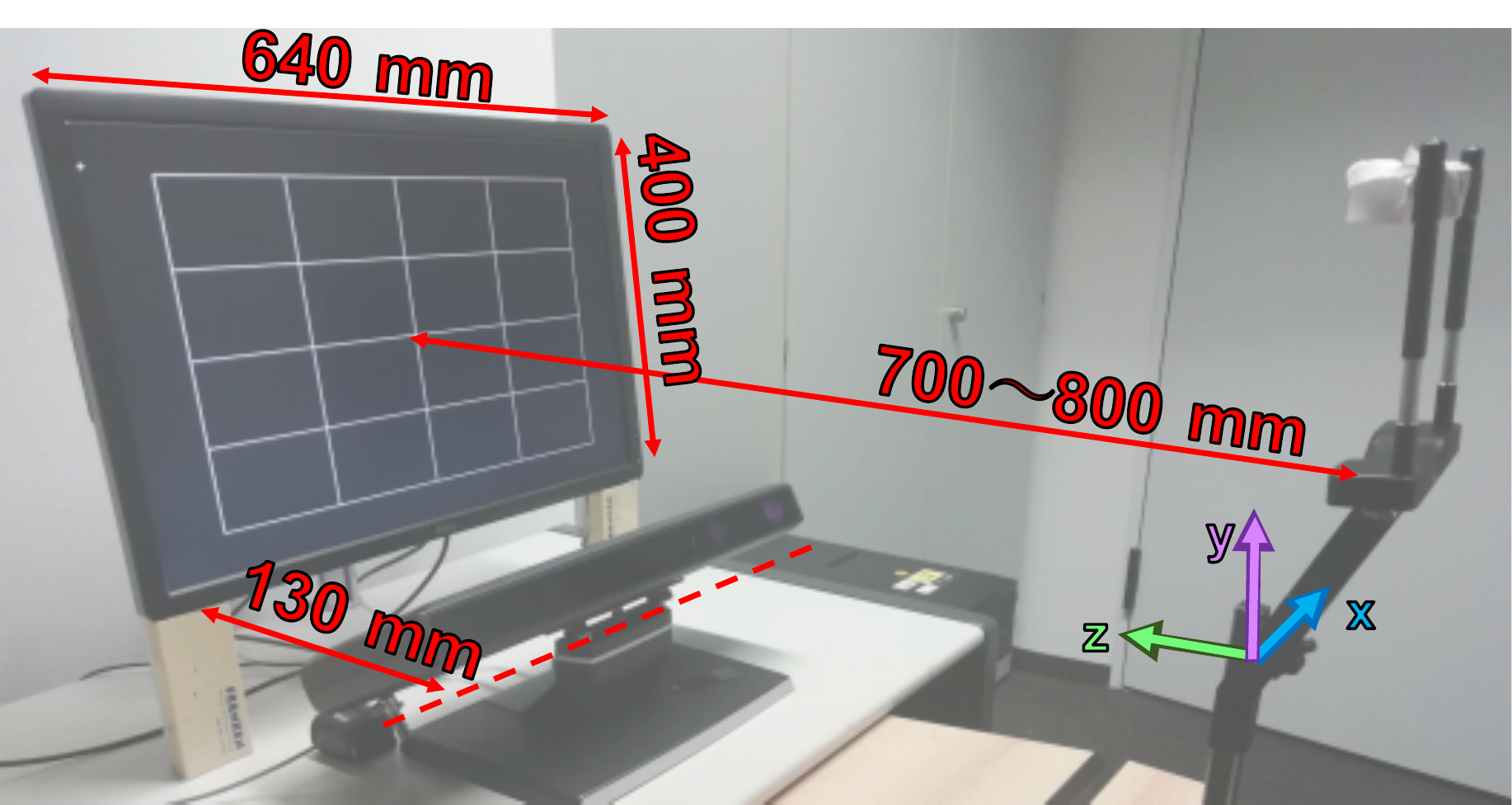}
    \caption{Experiment setup. A 30-inch monitor, a Tobii TX300 eye tracker and a chin rest were used for recording.  We adjusted the location of the table, where the chin rest was mounted on, across sessions to simulate head pose variation after initial calibration. The monitor size and relative device geometry are marked in the figure.}
    \label{fig:setup}
\end{figure}

% \begin{figure*}
%     \centering
%     \includegraphics[width=1.0\textwidth]{figures/grid2.pdf}
%     \caption{(a) shows the aggregated heat map of gaze points on a five-by-five grid over 10 participants in three different locations: the initial calibration location (left), a location with a small head pose change (middle) and another with a large head pose change (right). (b) presents three examples in the corresponding positions from one participant. The white grids indicate the ground truth gaze locations, while the red points and their connecting grids denote the gaze points and the distorted gaze projection plane. We see that the distortion of the gaze projection plane increases as head pose variation.}
%     \label{fig:grid}
% \end{figure*}

\subsection{Apparatus}

For data recording, we used a stationary Tobii TX300 eye tracker sampling at 300 Hz.
We used a chin rest to constrain head pose variation (see \autoref{fig:setup}).
The experiment interface was shown full-screen on a 30-inch (640x400 mm; $\sim$50$^{\circ}$, resolution of 2560x1600 pixels), which was placed approximate 750 mm away from the chin rest. 
The diameter of the circular stimuli before shrinking was 40 pixels ($\sim$0.8$^{\circ}$), corresponding to the best reported precision of the eye tracker.
%In our experiment, we instructed the participant to hold their head on a chin rest. 
We changed the chin rest location across sessions to maximize the inter-session differences caused by head pose and study such impact on our performance.
%However, the use of the chin rest was intended to constrain the head motion within each session to minimize the intra-session differences. 
This is because there should be an one-to-one mapping between head pose and the undistorted gaze projection plane, and the current study focuses on the plane correction without considering dynamic head pose changes.

\subsection{Participants}

We recruited 11 participants (three female; average age: 28.3), among which three are Indian, five are Asian, and three are Caucasian. 
Four of them wore glasses.
A close scrutiny reveals that the data of one participant had a poor condition, which presented an extra bad precision (i.e. a severer jittering/dispersion during fixation) and low accuracy (i.e. an obvious deviation from ground truth), and contained a large proportion of invalid tracking frames.
We therefore exclude this participant.
This gives us a 10-participant dataset.

\subsection{Procedure}

% \andreas{typically comes after apparatus and participants}

Participants were first introduced to the experiment interface and allowed to familiarize themselves with the interface for around a minute. 
Participants then performed a standard six-point calibration using the Tobii eye tracker interface, followed by three sessions of recordings, each of which lasted for about five minutes.
The first session directly followed the eye tracker calibration, while the second and the third sessions were conducted with a different amount of head position change.
This is to simulate the impact of head pose change after initial calibration. 
To this end, each time we adjusted the position of the chin rest randomly in x-, y-, and z-direction. 
% \andreas{add small coordinate system with xyz arrows in fig 6}
% in horizontal, vertical, and depth changes.
More specifically, the range of the first position change was in a medium degree (around 40 mm), and the second change was in a large degree (around 80 mm; reaching the boundary of the valid range of the eye tracker).
After each adjustment of the chin rest, participants were also allowed to tune the height and the position of the chair for the most comfortable condition.

In each session, the visual target traversed the five-by-five grid in a random order for five times, which resulted in around a hundred saccadic eye movements per participant.
In order words, our data in total contains 30 recording sessions and approximately 3,000 saccadic eye movements.
Between sessions participants were encouraged to rest, walk around, and look outside the window.
These breaks lasted for at least one minute and were extended up to around five minutes if requested by the participants.
On average, the entire experiment recording took about 20 minutes per participant.

\subsection{Distortion across  Pose Variations}

Two widely used metrics for eye tracking performance are precision and accuracy~\cite{duchowski2017eye,holmqvist2011eye}. 
\textit{Precision} represents the deviation among gaze points of one fixation from their centroid, while \textit{accuracy} denotes the average distance from gaze points to the ground truth location. 
Please note that our method aims to improve the eye tracking accuracy by transforming the deviated gaze points toward the correct locations. 
In other words, this is to amend the undermined accuracy (rather than precision) caused by the poor quality of initial calibration or the changes of screen-tracker geometry and head pose.

% \subsubsection{Obtaining gaze ground truth}
As we use the visual targets shown in a five-by-five grid to direct the saccadic eye movements, the ground truth location of the fixations was at the corresponding grid vertex.
To measure the accuracy, we calculate the average Euclidean distance between each grid vertex and the gaze points of all the fixations that correspond to the vertex over one session. 

We see that eye tracking accuracy generally decreases as the increase of head pose variation from  initial calibration position.
More specifically, the overall Euclidean distance between the gaze points and ground truth at  initial calibration pose is $1.07\pm0.65^{\circ}$, and those with small and large head pose variation are $1.17\pm0.64^{\circ}$ and $1.18\pm0.70^{\circ}$, respectively.

%% file: 06_evaluation.tex
\section{Experimental Evaluation}

% \andreas{the essential issues? ;)}

This section evaluates the effectiveness of the proposed eye-only calibration for reducing calibration distortion.
We aim to answer three key questions pertinent to this study: 
1) Can it improve eye tracking accuracy with and without head pose variation after initial calibration and across participants? 2) Is saccade selection necessary and what are the important attributes? 3) Which is the optimal projection method to correct distorted saccades?

In the following experimental evaluation, we present the results in a leave-one-participant-out paradigm. 
That is, each time we trained a random forest classifier on the data of nine participants and tested the undistortion effect on the left out participant. 
We repeated this process for 10 times and the final performance is the average result over all the iterations.

\subsection{Improvement over Initial Calibration}

\begin{figure*}
    \centering
    \begin{minipage}{.83\columnwidth}
        \centering
        \includegraphics[width=1.\columnwidth]{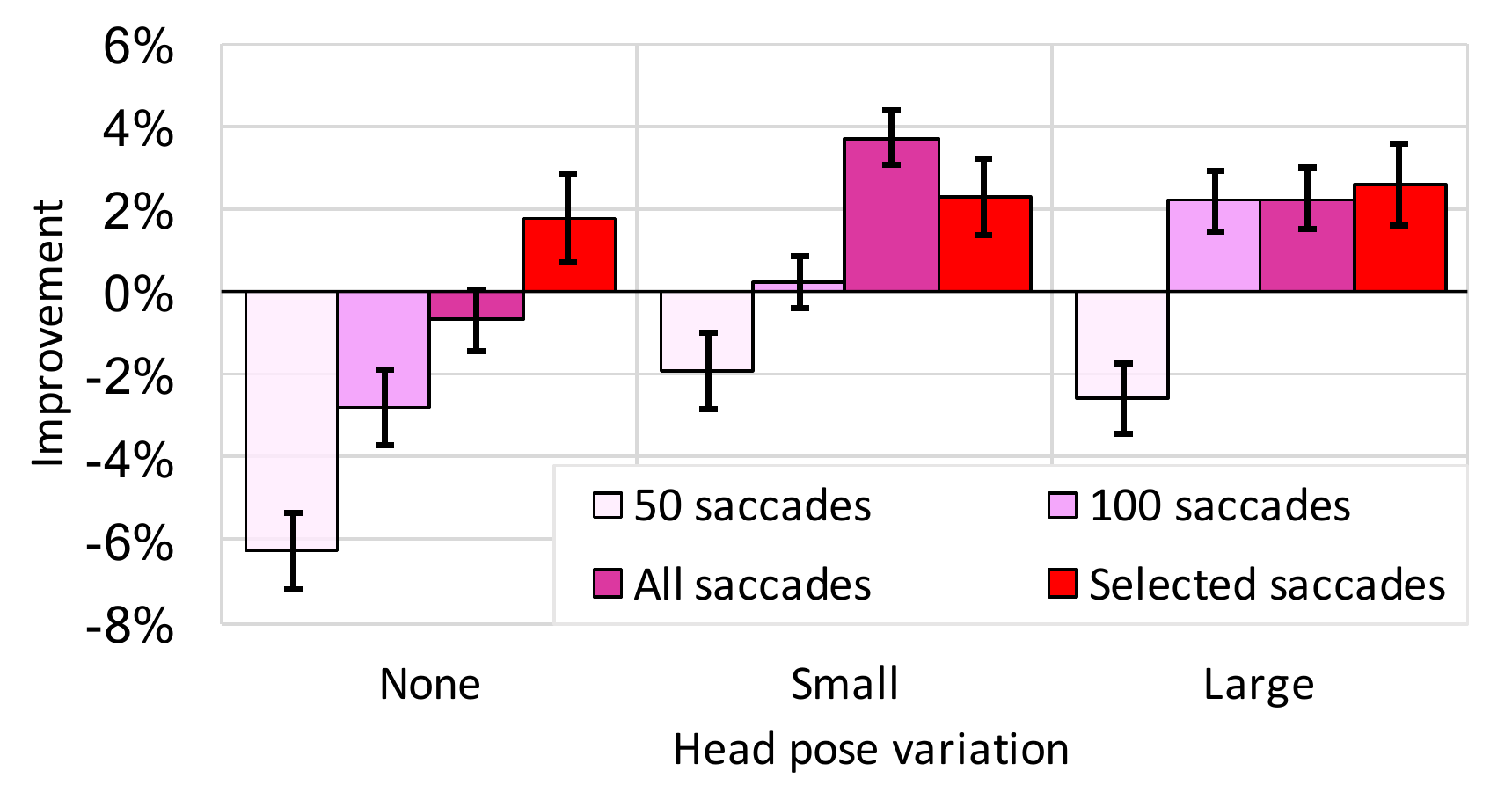} % first figure itself
        % \caption{Improvement over the initial calibration across different positions while using 50, 100, and all saccades and the data-driven selected saccades for plane correction. The error bars show the standard error. The data-driven saccade selection yields a stable improvement over all participants. The negative value indicates a decrease of accuracy.}
        % \label{fig:improvement_pos}
    \end{minipage}\hfill
    \begin{minipage}{.33\columnwidth}
        \centering
        \includegraphics[width=1.\columnwidth]{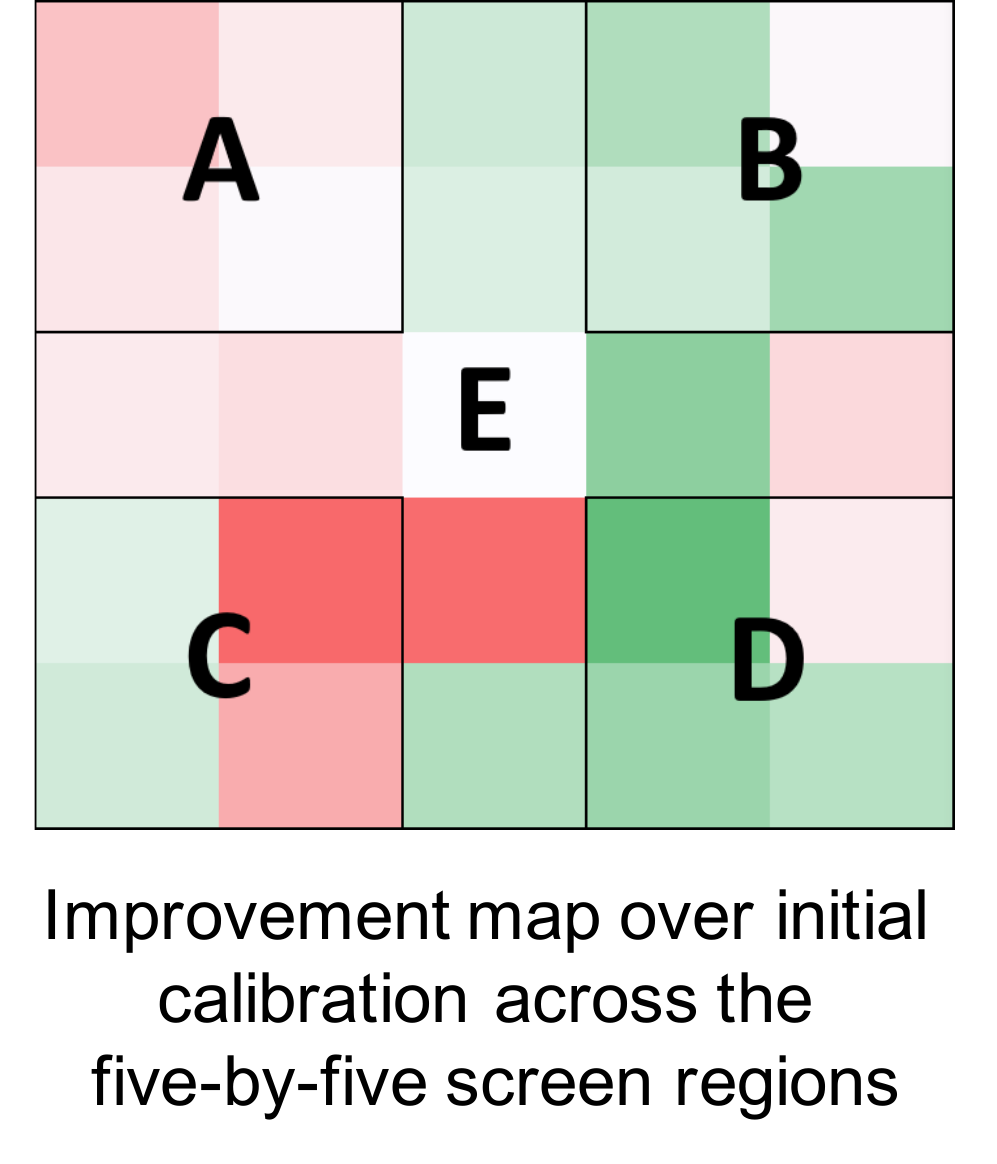} % first figure itself
        % \caption{lol.}
        % \label{fig:error_map}
    \end{minipage}\hfill
    \begin{minipage}{.83\columnwidth}
        \centering
        \includegraphics[width=1.\columnwidth]{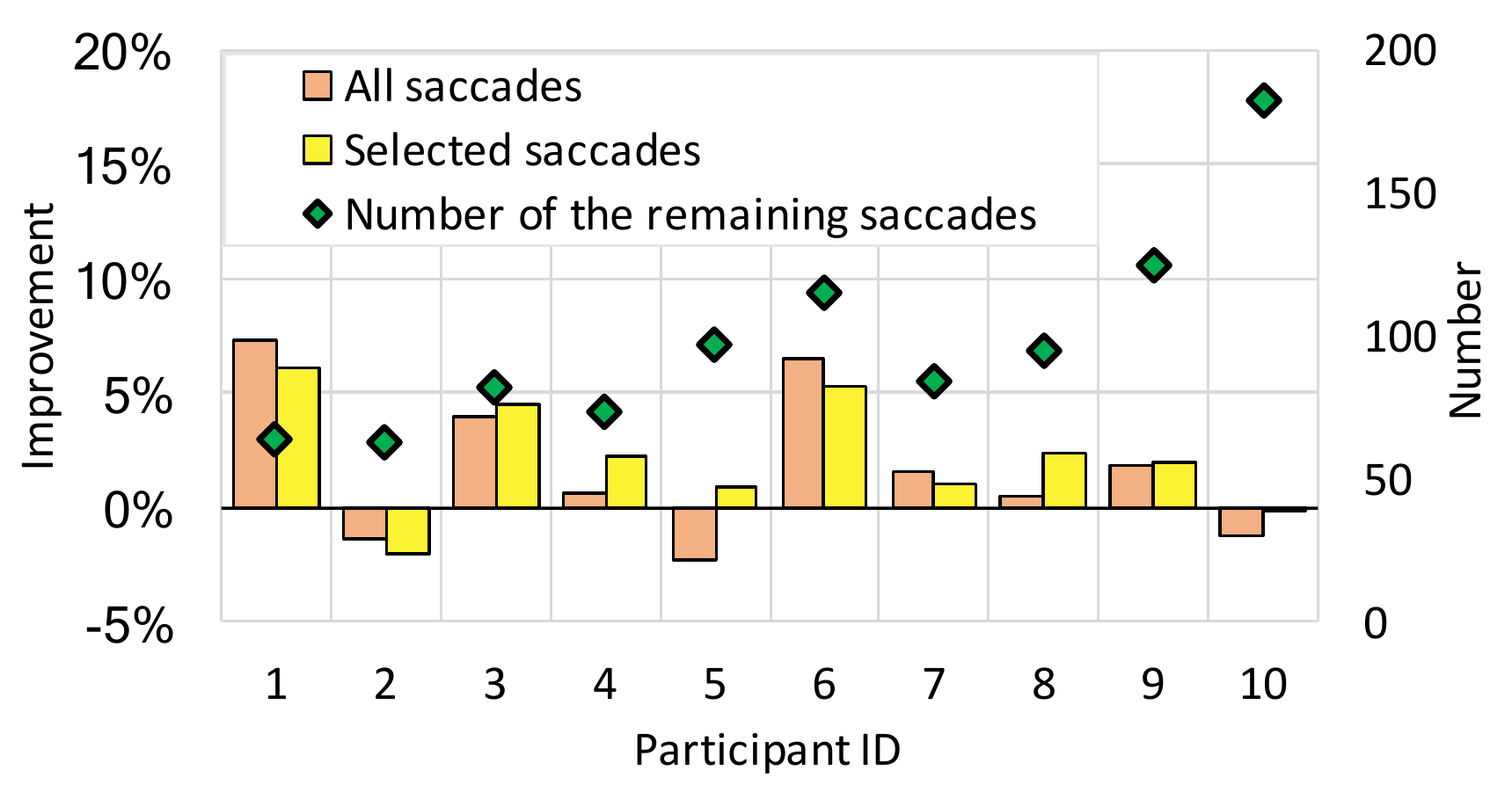} % second figure itself
        
    \end{minipage}
    \caption{(Left) Improvement compared to initial calibration for each participant. The bars indicate the improvement when using all vs selected saccades. The green diamonds show the number of remaining saccades from data-driven saccade selection. With this selection, our method achieves consistent improvements for most participants.
    (Middle) Improvement map over initial calibration. Green indicates positive while red negative. Improvement of our method varies across screen regions probably due to data skewness of suitable saccades.
    (Right) Improvement over initial calibration across three positions while using 50, 100, and all saccades and data-driven selected saccades for plane correction. The error bars show standard error. Data-driven saccade selection yields a stable improvement over all participants. The negative value indicates a accuracy decrease.}
        \label{fig:results}
\end{figure*}

\begin{figure*}[!b]
    \centering
    \includegraphics[width=1.\textwidth]{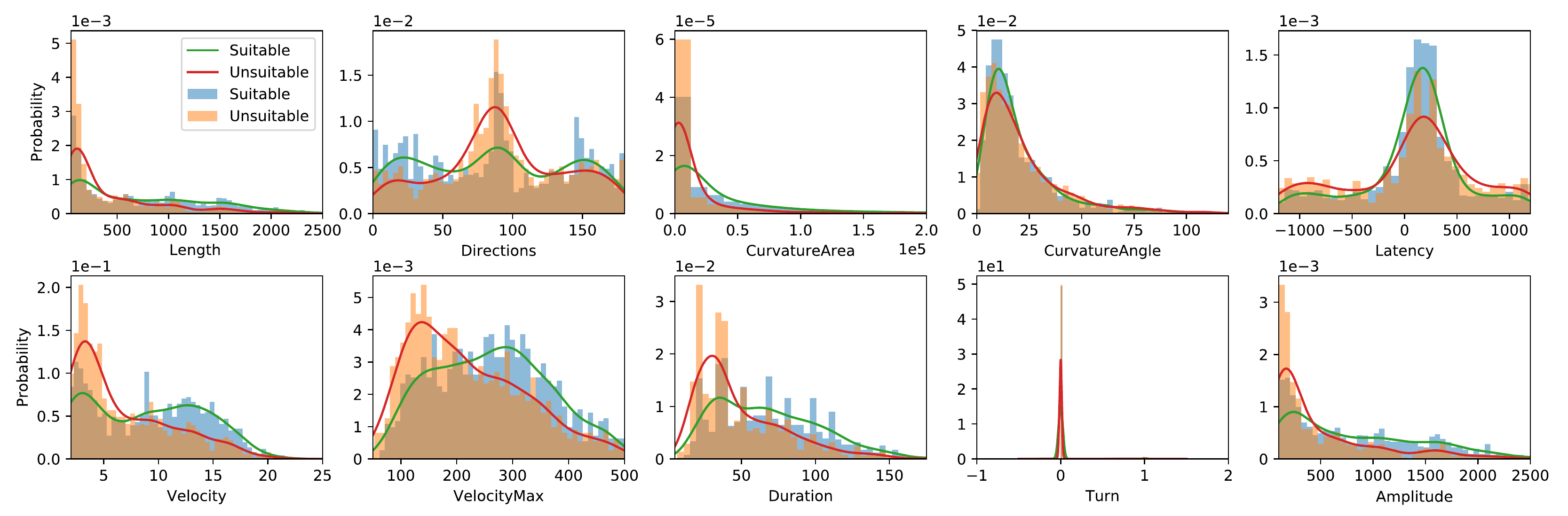}
    \caption{Probabilities mass functions of attributes of suitable and unsuitable saccades for the correction of gaze projection plane. Probability difference between suitable and unsuitable saccades exists in a certain value range of specific attributes.}
    \label{fig:attributes}
\end{figure*}

To answer the first question, we first look into the improvement over initial calibration of the eye tracker for overall participants. 
\autoref{fig:results} (left) shows the improvement across different head pose variations. 
Bars with different colors indicate plane correction based on different number (50, 100, and all) of saccades as well as saccades selected by data-driven approach (random forest). 
The black error bar presents standard error.  
For simplicity, we refer to initial calibration with none head pose variation as "None", head pose with a small variation as "Small" and with a large variation as "Large" in the following texts.
Most importantly, our method based on the data-driven saccade selection produces a consistent improvement over initial calibration with and without head pose variations. 
Overall, we achieve a 2.5\% improvement, pushing accuracy from $1.79^{\circ}$ to $1.75^{\circ}$ (equivalent to 36.3\% of average accuracy difference between "None" and "Large" without correction).

In contrast, undistorting gaze projection plane using 50 random saccades consistently decreases the accuracy, compared to that of using initial calibration.
Increasing the number of saccades to 100 improves the accuracy for "Small" and "Large", but still decreases in "None". 
Although further including all saccades (up to 300) improves the accuracy in both "None" and "Small" by a marked margin (2.1\%), it fails to improve the accuracy in "None" likewise.
This is probably because initial calibration without pose variation ($1.07\pm0.65^{\circ}$) is highly accurate and thus relatively hard to improve.

Interestingly, our improvement over initial calibration varies across screen regions (see~\autoref{fig:results} middle). 
Since we used a five-by-five grid of visual stimuli in data collection, we visualize the improvement map accordingly.
Green denotes accuracy increase and red decrease.
By and large, our method performs well to reduce distortion for overall participants, particularly on the right screen.
More specifically, our method can significantly improve over initial calibration for three-fourths of the screen area (B+C+D+E: p=0.028 or A+B+D+E: p=0.033).

Inspecting improvement on individual participant (see~\autoref{fig:results} right) suggests that our method can achieve improvement for majority (80\%) participants. 
Further, we can reach approximately 5\% improvement for almost one third of participants in overall scenarios across head pose variations.
Importantly, data-driven approach for saccade selection is beneficial to plane correction for 70\% participants. 
Interestingly, for a clear majority participants (70\%), the number of remaining saccades after  data-driven selection is between 50 to 100.
However, a random selection with a similar number of saccades fails to achieve equivalent accuracy (see \autoref{fig:results} Left), implying that correct saccade selection is essential to our method.

% \begin{figure}[b]
%     \centering
%     \includegraphics[width=1.0\columnwidth]{figures/importance.pdf}
%     \caption{Attribute importance of the random forest classifier for saccade selection.}
%     \label{fig:importance}
% \end{figure}

% \begin{figure}
%     \centering
%     \includegraphics[width=1.0\columnwidth]{figures/improvement_pos.pdf}
%     \caption{Improvement over the initial calibration across different positions while using 50, 100, and all saccades and the data-driven selected saccades for plane correction. The error bars show the standard error. The data-driven saccade selection yields a stable improvement over all participants. The negative value indicates a decrease of accuracy.
%     % \andreas{would ``None'' make more sense than ``initial'' (given that it's about head pose variation?); font sizes seem to differ; it's not easy to see what is good and what is bad here -- the positive or the negative values?}
%     }
%     \label{fig:improvement_pos}
% \end{figure}

% \begin{figure}[!t]
%     \centering
%     \includegraphics[width=1.0\columnwidth]{figures/improvement_user.pdf}
%     \caption{Improvement compared to the initial calibration for each participant. The bars indicate the improvement when using all vs selected saccades. The green diamonds show the number of the remaining saccades from the data-driven saccade selection. With this selection, our method achieves consistent improvements for most participants.}
%     \label{fig:improvement_user}
% \end{figure}

\subsection{Further Insight into Saccade Selection}

To understand the difference between suitable and unsuitable saccades for the correction of gaze projection plane, we plot the probability mass functions of each saccade attribute as shown in \autoref{fig:attributes}.
In general, the probabilities of suitable and unsuitable saccades have a large proportion of overlap.
However, their probabilities still peak at different values for some attributes, such as $Velocity$ and $VelocityMax$, suggesting that suitable saccades tend to have a higher velocity as well as a higher maximum velocity.
In addition, the chance of being a suitable or unsuitable saccade is relatively high in specific value range of some attributes.
For example, it is likely to be a suitable saccade with a large $Length$ or $Amplitude$, or a small positive $Latency$. 
In contrast, it is likely to be an unsuitable saccade with less than 50 ms duration.
That the probabilities of suitable and unsuitable saccades are overlapped, motivates us to apply the data-driven approach to identifying saccade candidates for plane correction.

% \autoref{fig:importance} shows the attribute importance analysis of our random forest classifiers through mean decrease impurity.
We also conducted the attribute importance analysis of our random forest classifiers through mean decrease impurity.
% Generally speaking, almost every attributes are useful, except for $Turn$, i.e. the number of big turn on a saccade.
% This is because $Turn$ has been used in the preprocessing of eye movement signals; the filtered saccades has no big turns by and large, with only a few exceptions with single turn. 
In good agreement with the previous discussion, $Length$ (ranks 2nd) and $Amplitude$ (3rd) rank high in attribute importance.
% , followed by $Velocity$. 
Most surprisingly, $CurvatureArea$ rank first, suggesting that $CurvatureArea$ can provide important complementary information for saccades selection, though itself alone is not informative enough (see \autoref{fig:attributes}).
This further indicates the need of data-driven saccade selection.

\subsection{Effect of Saccade Projection Methods}

\begin{figure}
    \centering
    \includegraphics[width=.90\columnwidth]{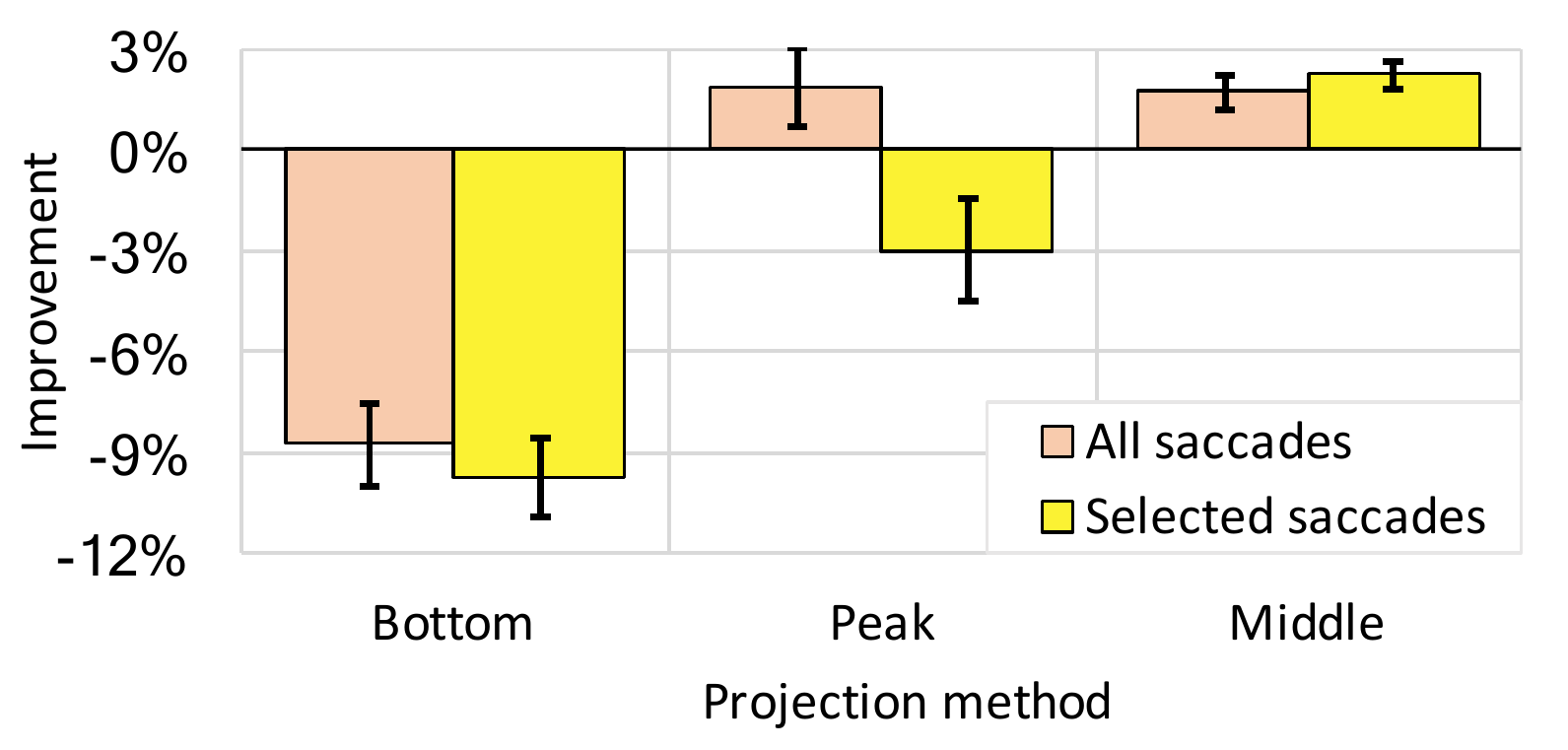}
    \caption{Performance comparison of different potential saccade projections. Projection to the middle between the saccade peak and bottom gives a consistent improvement over initial calibration using all and selected saccades.}
    \label{fig:projection}
\end{figure}

We discuss three different potential projections to correct the distorted saccades in the method section. 
We hereby evaluate the effectiveness of these projections. 
\autoref{fig:projection} shows the performance comparison of projecting the curve saccade to the line connecting two endpoints (Bottom) and its shifted counterpart to the saccade peak (Peak) and middle (Middle) between bottom and peak. 

In general, projecting saccades to the middle between the saccade peak and bottom as used in our method is most promising. 
It achieves improvements for both using all and selected saccades. 
In contrast, projecting to peak achieves improvement only with all saccades, while projecting to bottom decreases initial calibration accuracy regardless of using all or selected saccades.

Most importantly, this result suggests that projection to the middle is a stable solution to correcting the distortion of gaze projection plane.
It also points out the significant impact of different projection methods on the correction effect. 
As such, investigating alternative projection methods can be of great interest in future.

%% file: 07_discussion.tex
\section{Discussion}

In this work we presented the first eye-only calibration method to reduce calibration distortion without user input or expensive processing of on-screen content.
Specifically, we proposed to 
% reduce calibration drift by 
undistort gaze projection plane.
% by straightening the curved saccade trajectories.
% caused by a lack of eye tracker calibration and thus improve eye tracking accuracy.
As the first method of its kind, it was demonstrated to be effective. 
Further, we shed lights on two pertinent and critical issues: saccade selection and saccade projection method, i.e. what and how to undistort. 
These problems have been shown to be closely related to performance. 
As such, we believe this study represents a first important step towards eye-only calibration.
 % of eye trackers.

The results we achieved are encouraging.
%We achieve encouraging results in this novel study. 
The proposed method is able to improve eye tracking accuracy directly after initial calibration, where the accuracy is relatively high and difficult to improve.
Moreover, it can effectively reduce calibration distortion under small and large head pose variations after initial calibration. 
These improvements were consistent for most of our participants. 
For evaluation purposes of eye-only calibration, we collected a novel saccadic eye movement dataset.
%, where saccades have a higher chance of being straight by nature.
We believe the dataset will be beneficial to this new line of studies, and thus decided to release it upon acceptance and continue extending this dataset.

In practice, the proposed method has significant potential as a low-cost, non-intrusive, and privacy-preserving solution to reducing calibration distortion of stationary eye trackers.
First, our method does not rely on additional information, such as computational expensive saliency map. 
Second, it allows for implicit calibration while users naturally look at an interface.
Third, unlike previous approaches to eye tracker self-calibration, it does not require any additional user input or potentially privacy-sensitive information on on-screen content.
%, thus largely protecting the privacy sensitive user information. 
These properties are valuable for practical real-time gaze-based interfaces.

We also identified a number of interesting directions for future work.
First, due to the unbalance of ocular dominance~\cite{chaurasia1976eyedness}, the distortion of the gaze projection plane may differ for the left and the right eye.
While in the current study we only investigated the data of the left eye, it will be interesting 
%However, eye movements from two eye are somehow coupled. 
to study the relation between binocular coordination and the impact on the corresponding gaze projection planes. 
Second, in our experiment we ensured to capture saccades that are straight by nature by giving participants sufficient preparation time.
To improve practicality of the approach, future work could investigate saccade identification using microsaccades \cite{engbert2003microsaccades} or on how to ensure sufficient preparation time in user interface design.
% Third, we observed a few glissades, i.e.\ overshooting of saccadic eye movements. 
% Our method does not adapt to this behavior yet due to the small number of instances in our dataset.
% However, it may become an important issue for other experimental settings.
Finally, as the first work in this area of research, our paper lays important foundations for future work but there is, of course, room for general performance improvements of the method itself.
% \andreas{sentence on how this could be achived?}
For example, we plan to account for the resulting gaze projection plane in saccade selection, which can maintain a good on-screen distribution of the selected saccades and thus a better warping performance.

%% file: 08_conclusion.tex
\section{Conclusion}

In this work we proposed the first calibration method using saccadic eye movements that neither requires additional user input or expensive processing of on-screen content.
We demonstrated its potential in reducing calibration distortion on a new saccade dataset and compared its performance to initial calibration with and without head pose variations. 
As such, the method provides a low-cost, non-intrusive, and privacy-preserving solution to reduce calibration distortion.
We also identified two key challenges for eye-only calibration, namely saccade selection and saccade projection. 
While further research is required to make the approach practically usable, our results are promising and pave the way for a novel line of work on eye-only calibration for stationary eye trackers.

% terms:
% undistort the plane
% undistort the saccade or project/ correct?
% distorted and undistorted gaze projection plane
% curved and straight saccade
% gaze points (no estimate)
% training samples (i.e. saccade)

%% file: 00_main.bbl
%%% -*-BibTeX-*-
%%% Do NOT edit. File created by BibTeX with style
%%% ACM-Reference-Format-Journals [18-Jan-2012].

\begin{thebibliography}{59}

%%% ====================================================================
%%% NOTE TO THE USER: you can override these defaults by providing
%%% customized versions of any of these macros before the \bibliography
%%% command.  Each of them MUST provide its own final punctuation,
%%% except for \shownote{}, \showDOI{}, and \showURL{}.  The latter two
%%% do not use final punctuation, in order to avoid confusing it with
%%% the Web address.
%%%
%%% To suppress output of a particular field, define its macro to expand
%%% to an empty string, or better, \unskip, like this:
%%%
%%% \newcommand{\showDOI}[1]{\unskip}   % LaTeX syntax
%%%
%%% \def \showDOI #1{\unskip}           % plain TeX syntax
%%%
%%% ====================================================================

\ifx \showCODEN    \undefined \def \showCODEN     #1{\unskip}     \fi
\ifx \showDOI      \undefined \def \showDOI       #1{#1}\fi
\ifx \showISBNx    \undefined \def \showISBNx     #1{\unskip}     \fi
\ifx \showISBNxiii \undefined \def \showISBNxiii  #1{\unskip}     \fi
\ifx \showISSN     \undefined \def \showISSN      #1{\unskip}     \fi
\ifx \showLCCN     \undefined \def \showLCCN      #1{\unskip}     \fi
\ifx \shownote     \undefined \def \shownote      #1{#1}          \fi
\ifx \showarticletitle \undefined \def \showarticletitle #1{#1}   \fi
\ifx \showURL      \undefined \def \showURL       {\relax}        \fi
% The following commands are used for tagged output and should be
% invisible to TeX
\providecommand\bibfield[2]{#2}
\providecommand\bibinfo[2]{#2}
\providecommand\natexlab[1]{#1}
\providecommand\showeprint[2][]{arXiv:#2}

\bibitem[\protect\citeauthoryear{Arabadzhiyska, Tursun, Myszkowski, Seidel, and
  Didyk}{Arabadzhiyska et~al\mbox{.}}{2017}]%
        {arabadzhiyska2017saccade}
\bibfield{author}{\bibinfo{person}{Elena Arabadzhiyska},
  \bibinfo{person}{Okan~Tarhan Tursun}, \bibinfo{person}{Karol Myszkowski},
  \bibinfo{person}{Hans-Peter Seidel}, {and} \bibinfo{person}{Piotr Didyk}.}
  \bibinfo{year}{2017}\natexlab{}.
\newblock \showarticletitle{Saccade landing position prediction for
  gaze-contingent rendering}.
\newblock \bibinfo{journal}{\emph{ACM Transactions on Graphics (TOG)}}
  \bibinfo{volume}{36}, \bibinfo{number}{4} (\bibinfo{year}{2017}),
  \bibinfo{pages}{50}.
\newblock


\bibitem[\protect\citeauthoryear{Bahill and Stark}{Bahill and Stark}{1975}]%
        {bahill1975neurological}
\bibfield{author}{\bibinfo{person}{A~Terry Bahill} {and}
  \bibinfo{person}{Lawrence Stark}.} \bibinfo{year}{1975}\natexlab{}.
\newblock \showarticletitle{Neurological control of horizontal and vertical
  components of oblique saccadic eye movements}.
\newblock \bibinfo{journal}{\emph{Mathematical Biosciences}}
  \bibinfo{volume}{27}, \bibinfo{number}{3-4} (\bibinfo{year}{1975}),
  \bibinfo{pages}{287--298}.
\newblock


\bibitem[\protect\citeauthoryear{Barz, Daiber, Sonntag, and Bulling}{Barz
  et~al\mbox{.}}{2018}]%
        {barz18_etra}
\bibfield{author}{\bibinfo{person}{Michael Barz}, \bibinfo{person}{Florian
  Daiber}, \bibinfo{person}{Daniel Sonntag}, {and} \bibinfo{person}{Andreas
  Bulling}.} \bibinfo{year}{2018}\natexlab{}.
\newblock \showarticletitle{Error-Aware Gaze-Based Interfaces for Robust Mobile
  Gaze Interaction}. In \bibinfo{booktitle}{\emph{Proc. International Symposium
  on Eye Tracking Research and Applications (ETRA)}}.
  \bibinfo{pages}{24:1--24:10}.
\newblock
\urldef\tempurl%
\url{https://doi.org/10.1145/3204493.3204536}
\showDOI{\tempurl}


\bibitem[\protect\citeauthoryear{Blignaut}{Blignaut}{2016}]%
        {blignaut2016idiosyncratic}
\bibfield{author}{\bibinfo{person}{Pieter Blignaut}.}
  \bibinfo{year}{2016}\natexlab{}.
\newblock \showarticletitle{Idiosyncratic feature-based gaze mapping}.
\newblock \bibinfo{journal}{\emph{JOURNAL OF EYE MOVEMENT RESEARCH}}
  \bibinfo{volume}{9}, \bibinfo{number}{3} (\bibinfo{year}{2016}).
\newblock


\bibitem[\protect\citeauthoryear{Boghen, Troost, Daroff, Dell'Osso, and
  Birkett}{Boghen et~al\mbox{.}}{1974}]%
        {boghen1974velocity}
\bibfield{author}{\bibinfo{person}{D Boghen}, \bibinfo{person}{BT Troost},
  \bibinfo{person}{RB Daroff}, \bibinfo{person}{LF Dell'Osso}, {and}
  \bibinfo{person}{JE Birkett}.} \bibinfo{year}{1974}\natexlab{}.
\newblock \showarticletitle{Velocity characteristics of normal human saccades}.
\newblock \bibinfo{journal}{\emph{Investigative Ophthalmology \& Visual
  Science}} \bibinfo{volume}{13}, \bibinfo{number}{8} (\bibinfo{year}{1974}),
  \bibinfo{pages}{619--623}.
\newblock


\bibitem[\protect\citeauthoryear{Borji, Sihite, and Itti}{Borji
  et~al\mbox{.}}{2012}]%
        {borji2012probabilistic}
\bibfield{author}{\bibinfo{person}{Ali Borji}, \bibinfo{person}{Dicky~N
  Sihite}, {and} \bibinfo{person}{Laurent Itti}.}
  \bibinfo{year}{2012}\natexlab{}.
\newblock \showarticletitle{Probabilistic learning of task-specific visual
  attention}. In \bibinfo{booktitle}{\emph{Computer Vision and Pattern
  Recognition (CVPR), 2012 IEEE Conference on}}. IEEE,
  \bibinfo{pages}{470--477}.
\newblock


\bibitem[\protect\citeauthoryear{Chaurasia and Mathur}{Chaurasia and
  Mathur}{1976}]%
        {chaurasia1976eyedness}
\bibfield{author}{\bibinfo{person}{BD Chaurasia} {and} \bibinfo{person}{BBL
  Mathur}.} \bibinfo{year}{1976}\natexlab{}.
\newblock \showarticletitle{Eyedness}.
\newblock \bibinfo{journal}{\emph{Cells Tissues Organs}} \bibinfo{volume}{96},
  \bibinfo{number}{2} (\bibinfo{year}{1976}), \bibinfo{pages}{301--305}.
\newblock


\bibitem[\protect\citeauthoryear{Chen and Ji}{Chen and Ji}{2015}]%
        {chen2015probabilistic}
\bibfield{author}{\bibinfo{person}{Jixu Chen} {and} \bibinfo{person}{Qiang
  Ji}.} \bibinfo{year}{2015}\natexlab{}.
\newblock \showarticletitle{A probabilistic approach to online eye gaze
  tracking without explicit personal calibration}.
\newblock \bibinfo{journal}{\emph{IEEE Transactions on Image Processing}}
  \bibinfo{volume}{24}, \bibinfo{number}{3} (\bibinfo{year}{2015}),
  \bibinfo{pages}{1076--1086}.
\newblock


\bibitem[\protect\citeauthoryear{Dorr, Martinetz, Gegenfurtner, and Barth}{Dorr
  et~al\mbox{.}}{2010}]%
        {dorr2010variability}
\bibfield{author}{\bibinfo{person}{Michael Dorr}, \bibinfo{person}{Thomas
  Martinetz}, \bibinfo{person}{Karl~R Gegenfurtner}, {and}
  \bibinfo{person}{Erhardt Barth}.} \bibinfo{year}{2010}\natexlab{}.
\newblock \showarticletitle{Variability of eye movements when viewing dynamic
  natural scenes}.
\newblock \bibinfo{journal}{\emph{Journal of vision}} \bibinfo{volume}{10},
  \bibinfo{number}{10} (\bibinfo{year}{2010}), \bibinfo{pages}{28--28}.
\newblock


\bibitem[\protect\citeauthoryear{Doyle and Walker}{Doyle and Walker}{2001}]%
        {doyle2001curved}
\bibfield{author}{\bibinfo{person}{Melanie Doyle} {and} \bibinfo{person}{Robin
  Walker}.} \bibinfo{year}{2001}\natexlab{}.
\newblock \showarticletitle{Curved saccade trajectories: Voluntary and
  reflexive saccades curve away from irrelevant distractors}.
\newblock \bibinfo{journal}{\emph{Experimental Brain Research}}
  \bibinfo{volume}{139}, \bibinfo{number}{3} (\bibinfo{year}{2001}),
  \bibinfo{pages}{333--344}.
\newblock


\bibitem[\protect\citeauthoryear{Duchowski}{Duchowski}{2017}]%
        {duchowski2017eye}
\bibfield{author}{\bibinfo{person}{Andrew~T Duchowski}.}
  \bibinfo{year}{2017}\natexlab{}.
\newblock \showarticletitle{Eye Tracking Methodology: Theory and Practice}.
\newblock  (\bibinfo{year}{2017}).
\newblock


\bibitem[\protect\citeauthoryear{Engbert and Kliegl}{Engbert and
  Kliegl}{2003}]%
        {engbert2003microsaccades}
\bibfield{author}{\bibinfo{person}{Ralf Engbert} {and}
  \bibinfo{person}{Reinhold Kliegl}.} \bibinfo{year}{2003}\natexlab{}.
\newblock \showarticletitle{Microsaccades uncover the orientation of covert
  attention}.
\newblock \bibinfo{journal}{\emph{Vision research}} \bibinfo{volume}{43},
  \bibinfo{number}{9} (\bibinfo{year}{2003}), \bibinfo{pages}{1035--1045}.
\newblock


\bibitem[\protect\citeauthoryear{Fan, Chen, Wei, Wang, and Zhu}{Fan
  et~al\mbox{.}}{2018}]%
        {fan2018inferring}
\bibfield{author}{\bibinfo{person}{Lifeng Fan}, \bibinfo{person}{Yixin Chen},
  \bibinfo{person}{Ping Wei}, \bibinfo{person}{Wenguan Wang}, {and}
  \bibinfo{person}{Song-Chun Zhu}.} \bibinfo{year}{2018}\natexlab{}.
\newblock \showarticletitle{Inferring Shared Attention in Social Scene Videos}.
  In \bibinfo{booktitle}{\emph{Proceedings of the IEEE Conference on Computer
  Vision and Pattern Recognition}}. \bibinfo{pages}{6460--6468}.
\newblock


\bibitem[\protect\citeauthoryear{Farmer and Sidorowich}{Farmer and
  Sidorowich}{1991}]%
        {farmer1991optimal}
\bibfield{author}{\bibinfo{person}{J~Doyne Farmer} {and}
  \bibinfo{person}{John~J Sidorowich}.} \bibinfo{year}{1991}\natexlab{}.
\newblock \showarticletitle{Optimal shadowing and noise reduction}.
\newblock \bibinfo{journal}{\emph{Physica D: Nonlinear Phenomena}}
  \bibinfo{volume}{47}, \bibinfo{number}{3} (\bibinfo{year}{1991}),
  \bibinfo{pages}{373--392}.
\newblock


\bibitem[\protect\citeauthoryear{Godijn and Theeuwes}{Godijn and
  Theeuwes}{2004}]%
        {godijn2004relationship}
\bibfield{author}{\bibinfo{person}{Richard Godijn} {and} \bibinfo{person}{Jan
  Theeuwes}.} \bibinfo{year}{2004}\natexlab{}.
\newblock \showarticletitle{The relationship between inhibition of return and
  saccade trajectory deviations.}
\newblock \bibinfo{journal}{\emph{Journal of Experimental Psychology: Human
  Perception and Performance}} \bibinfo{volume}{30}, \bibinfo{number}{3}
  (\bibinfo{year}{2004}), \bibinfo{pages}{538}.
\newblock


\bibitem[\protect\citeauthoryear{Gorji and Clark}{Gorji and Clark}{2017}]%
        {gorji2017attentional}
\bibfield{author}{\bibinfo{person}{Siavash Gorji} {and}
  \bibinfo{person}{James~J Clark}.} \bibinfo{year}{2017}\natexlab{}.
\newblock \showarticletitle{Attentional push: A deep convolutional network for
  augmenting image salience with shared attention modeling in social scenes}.
  In \bibinfo{booktitle}{\emph{Computer Vision and Pattern Recognition (CVPR),
  2017 IEEE Conference on}}, Vol.~\bibinfo{volume}{2}. IEEE,
  \bibinfo{pages}{5}.
\newblock


\bibitem[\protect\citeauthoryear{Holmqvist, Nystr{\"o}m, Andersson, Dewhurst,
  Jarodzka, and Van~de Weijer}{Holmqvist et~al\mbox{.}}{2011}]%
        {holmqvist2011eye}
\bibfield{author}{\bibinfo{person}{Kenneth Holmqvist}, \bibinfo{person}{Marcus
  Nystr{\"o}m}, \bibinfo{person}{Richard Andersson}, \bibinfo{person}{Richard
  Dewhurst}, \bibinfo{person}{Halszka Jarodzka}, {and} \bibinfo{person}{Joost
  Van~de Weijer}.} \bibinfo{year}{2011}\natexlab{}.
\newblock \bibinfo{booktitle}{\emph{Eye tracking: A comprehensive guide to
  methods and measures}}.
\newblock \bibinfo{publisher}{OUP Oxford}.
\newblock


\bibitem[\protect\citeauthoryear{Huang, White, and Buscher}{Huang
  et~al\mbox{.}}{2012}]%
        {huang2012user}
\bibfield{author}{\bibinfo{person}{Jeff Huang}, \bibinfo{person}{Ryen White},
  {and} \bibinfo{person}{Georg Buscher}.} \bibinfo{year}{2012}\natexlab{}.
\newblock \showarticletitle{User see, user point: gaze and cursor alignment in
  web search}. In \bibinfo{booktitle}{\emph{Proceedings of the SIGCHI
  Conference on Human Factors in Computing Systems}}. ACM,
  \bibinfo{pages}{1341--1350}.
\newblock


\bibitem[\protect\citeauthoryear{Huang, Kwok, Ngai, Chan, and Leong}{Huang
  et~al\mbox{.}}{2016}]%
        {huang2016building}
\bibfield{author}{\bibinfo{person}{Michael~Xuelin Huang},
  \bibinfo{person}{Tiffany~CK Kwok}, \bibinfo{person}{Grace Ngai},
  \bibinfo{person}{Stephen~CF Chan}, {and} \bibinfo{person}{Hong~Va Leong}.}
  \bibinfo{year}{2016}\natexlab{}.
\newblock \showarticletitle{Building a personalized, auto-calibrating eye
  tracker from user interactions}. In \bibinfo{booktitle}{\emph{Proceedings of
  the 2016 CHI Conference on Human Factors in Computing Systems}}. ACM,
  \bibinfo{pages}{5169--5179}.
\newblock


\bibitem[\protect\citeauthoryear{Huang, Li, Ngai, and Leong}{Huang
  et~al\mbox{.}}{2017}]%
        {huang2017screenglint}
\bibfield{author}{\bibinfo{person}{Michael~Xuelin Huang},
  \bibinfo{person}{Jiajia Li}, \bibinfo{person}{Grace Ngai}, {and}
  \bibinfo{person}{Hong~Va Leong}.} \bibinfo{year}{2017}\natexlab{}.
\newblock \showarticletitle{Screenglint: Practical, in-situ gaze estimation on
  smartphones}. In \bibinfo{booktitle}{\emph{Proceedings of the 2017 CHI
  Conference on Human Factors in Computing Systems}}. ACM,
  \bibinfo{pages}{2546--2557}.
\newblock


\bibitem[\protect\citeauthoryear{Huang, Shen, Boix, and Zhao}{Huang
  et~al\mbox{.}}{2015}]%
        {huang2015salicon}
\bibfield{author}{\bibinfo{person}{Xun Huang}, \bibinfo{person}{Chengyao Shen},
  \bibinfo{person}{Xavier Boix}, {and} \bibinfo{person}{Qi Zhao}.}
  \bibinfo{year}{2015}\natexlab{}.
\newblock \showarticletitle{Salicon: Reducing the semantic gap in saliency
  prediction by adapting deep neural networks}. In
  \bibinfo{booktitle}{\emph{Proceedings of the IEEE International Conference on
  Computer Vision}}. \bibinfo{pages}{262--270}.
\newblock


\bibitem[\protect\citeauthoryear{Huang, Cai, Li, and Sato}{Huang
  et~al\mbox{.}}{2018}]%
        {huang2018predicting}
\bibfield{author}{\bibinfo{person}{Yifei Huang}, \bibinfo{person}{Minjie Cai},
  \bibinfo{person}{Zhenqiang Li}, {and} \bibinfo{person}{Yoichi Sato}.}
  \bibinfo{year}{2018}\natexlab{}.
\newblock \showarticletitle{Predicting Gaze in Egocentric Video by Learning
  Task-dependent Attention Transition}. In \bibinfo{booktitle}{\emph{European
  Conference on Computer Vision}}.
\newblock


\bibitem[\protect\citeauthoryear{Itti, Koch, and Niebur}{Itti
  et~al\mbox{.}}{1998}]%
        {itti1998model}
\bibfield{author}{\bibinfo{person}{Laurent Itti}, \bibinfo{person}{Christof
  Koch}, {and} \bibinfo{person}{Ernst Niebur}.}
  \bibinfo{year}{1998}\natexlab{}.
\newblock \showarticletitle{A model of saliency-based visual attention for
  rapid scene analysis}.
\newblock \bibinfo{journal}{\emph{IEEE Transactions on pattern analysis and
  machine intelligence}} \bibinfo{volume}{20}, \bibinfo{number}{11}
  (\bibinfo{year}{1998}), \bibinfo{pages}{1254--1259}.
\newblock


\bibitem[\protect\citeauthoryear{Judd, Ehinger, Durand, and Torralba}{Judd
  et~al\mbox{.}}{2009}]%
        {judd2009learning}
\bibfield{author}{\bibinfo{person}{Tilke Judd}, \bibinfo{person}{Krista
  Ehinger}, \bibinfo{person}{Fr{\'e}do Durand}, {and} \bibinfo{person}{Antonio
  Torralba}.} \bibinfo{year}{2009}\natexlab{}.
\newblock \showarticletitle{Learning to predict where humans look}. In
  \bibinfo{booktitle}{\emph{Computer Vision, 2009 IEEE 12th international
  conference on}}. IEEE, \bibinfo{pages}{2106--2113}.
\newblock


\bibitem[\protect\citeauthoryear{Khamis, Saltuk, Hang, Stolz, Bulling, and
  Alt}{Khamis et~al\mbox{.}}{2016}]%
        {khamis2016textpursuits}
\bibfield{author}{\bibinfo{person}{Mohamed Khamis}, \bibinfo{person}{Ozan
  Saltuk}, \bibinfo{person}{Alina Hang}, \bibinfo{person}{Katharina Stolz},
  \bibinfo{person}{Andreas Bulling}, {and} \bibinfo{person}{Florian Alt}.}
  \bibinfo{year}{2016}\natexlab{}.
\newblock \showarticletitle{TextPursuits: using text for pursuits-based
  interaction and calibration on public displays}. In
  \bibinfo{booktitle}{\emph{Proceedings of the 2016 ACM International Joint
  Conference on Pervasive and Ubiquitous Computing}}. ACM,
  \bibinfo{pages}{274--285}.
\newblock


\bibitem[\protect\citeauthoryear{Koch and Ullman}{Koch and Ullman}{1987}]%
        {koch1987shifts}
\bibfield{author}{\bibinfo{person}{Christof Koch} {and} \bibinfo{person}{Shimon
  Ullman}.} \bibinfo{year}{1987}\natexlab{}.
\newblock \showarticletitle{Shifts in selective visual attention: towards the
  underlying neural circuitry}.
\newblock In \bibinfo{booktitle}{\emph{Matters of intelligence}}.
  \bibinfo{publisher}{Springer}, \bibinfo{pages}{115--141}.
\newblock


\bibitem[\protect\citeauthoryear{Kowler}{Kowler}{2011}]%
        {kowler2011eye}
\bibfield{author}{\bibinfo{person}{Eileen Kowler}.}
  \bibinfo{year}{2011}\natexlab{}.
\newblock \showarticletitle{Eye movements: The past 25 years}.
\newblock \bibinfo{journal}{\emph{Vision research}} \bibinfo{volume}{51},
  \bibinfo{number}{13} (\bibinfo{year}{2011}), \bibinfo{pages}{1457--1483}.
\newblock


\bibitem[\protect\citeauthoryear{Kruijne, Van~der Stigchel, and Meeter}{Kruijne
  et~al\mbox{.}}{2014}]%
        {kruijne2014model}
\bibfield{author}{\bibinfo{person}{Wouter Kruijne}, \bibinfo{person}{Stefan
  Van~der Stigchel}, {and} \bibinfo{person}{Martijn Meeter}.}
  \bibinfo{year}{2014}\natexlab{}.
\newblock \showarticletitle{A model of curved saccade trajectories: Spike rate
  adaptation in the brainstem as the cause of deviation away}.
\newblock \bibinfo{journal}{\emph{Brain and cognition}}  \bibinfo{volume}{85}
  (\bibinfo{year}{2014}), \bibinfo{pages}{259--270}.
\newblock


\bibitem[\protect\citeauthoryear{Ludwig and Gilchrist}{Ludwig and
  Gilchrist}{2003}]%
        {ludwig2003target}
\bibfield{author}{\bibinfo{person}{Casimir~JH Ludwig} {and}
  \bibinfo{person}{Iain~D Gilchrist}.} \bibinfo{year}{2003}\natexlab{}.
\newblock \showarticletitle{Target similarity affects saccade curvature away
  from irrelevant onsets}.
\newblock \bibinfo{journal}{\emph{Experimental Brain Research}}
  \bibinfo{volume}{152}, \bibinfo{number}{1} (\bibinfo{year}{2003}),
  \bibinfo{pages}{60--69}.
\newblock


\bibitem[\protect\citeauthoryear{McPeek, Han, and Keller}{McPeek
  et~al\mbox{.}}{2003}]%
        {mcpeek2003competition}
\bibfield{author}{\bibinfo{person}{Robert~M McPeek}, \bibinfo{person}{Jae~H
  Han}, {and} \bibinfo{person}{Edward~L Keller}.}
  \bibinfo{year}{2003}\natexlab{}.
\newblock \showarticletitle{Competition between saccade goals in the superior
  colliculus produces saccade curvature}.
\newblock \bibinfo{journal}{\emph{Journal of Neurophysiology}}
  \bibinfo{volume}{89}, \bibinfo{number}{5} (\bibinfo{year}{2003}),
  \bibinfo{pages}{2577--2590}.
\newblock


\bibitem[\protect\citeauthoryear{McSorley, Haggard, and Walker}{McSorley
  et~al\mbox{.}}{2006}]%
        {mcsorley2006time}
\bibfield{author}{\bibinfo{person}{Eugene McSorley}, \bibinfo{person}{Patrick
  Haggard}, {and} \bibinfo{person}{Robin Walker}.}
  \bibinfo{year}{2006}\natexlab{}.
\newblock \showarticletitle{Time course of oculomotor inhibition revealed by
  saccade trajectory modulation}.
\newblock \bibinfo{journal}{\emph{Journal of Neurophysiology}}
  \bibinfo{volume}{96}, \bibinfo{number}{3} (\bibinfo{year}{2006}),
  \bibinfo{pages}{1420--1424}.
\newblock


\bibitem[\protect\citeauthoryear{Megardon, Ludwig, and Sumner}{Megardon
  et~al\mbox{.}}{2017}]%
        {megardon2017trajectory}
\bibfield{author}{\bibinfo{person}{Geoffrey Megardon}, \bibinfo{person}{Casimir
  Ludwig}, {and} \bibinfo{person}{Petroc Sumner}.}
  \bibinfo{year}{2017}\natexlab{}.
\newblock \showarticletitle{Trajectory curvature in saccade sequences:
  spatiotopic influences vs. residual motor activity}.
\newblock \bibinfo{journal}{\emph{Journal of Neurophysiology}}
  \bibinfo{volume}{118}, \bibinfo{number}{2} (\bibinfo{year}{2017}),
  \bibinfo{pages}{1310--1320}.
\newblock


\bibitem[\protect\citeauthoryear{Moehler and Fiehler}{Moehler and
  Fiehler}{2014}]%
        {moehler2014effects}
\bibfield{author}{\bibinfo{person}{Tobias Moehler} {and} \bibinfo{person}{Katja
  Fiehler}.} \bibinfo{year}{2014}\natexlab{}.
\newblock \showarticletitle{Effects of spatial congruency on saccade and visual
  discrimination performance in a dual-task paradigm}.
\newblock \bibinfo{journal}{\emph{Vision research}}  \bibinfo{volume}{105}
  (\bibinfo{year}{2014}), \bibinfo{pages}{100--111}.
\newblock


\bibitem[\protect\citeauthoryear{Moehler and Fiehler}{Moehler and
  Fiehler}{2015}]%
        {moehler2015influence}
\bibfield{author}{\bibinfo{person}{Tobias Moehler} {and} \bibinfo{person}{Katja
  Fiehler}.} \bibinfo{year}{2015}\natexlab{}.
\newblock \showarticletitle{The influence of spatial congruency and movement
  preparation time on saccade curvature in simultaneous and sequential
  dual-tasks}.
\newblock \bibinfo{journal}{\emph{Vision research}}  \bibinfo{volume}{116}
  (\bibinfo{year}{2015}), \bibinfo{pages}{25--35}.
\newblock


\bibitem[\protect\citeauthoryear{Papoutsaki, Sangkloy, Laskey, Daskalova,
  Huang, and Hays}{Papoutsaki et~al\mbox{.}}{2016}]%
        {papoutsaki2016webgazer}
\bibfield{author}{\bibinfo{person}{Alexandra Papoutsaki},
  \bibinfo{person}{Patsorn Sangkloy}, \bibinfo{person}{James Laskey},
  \bibinfo{person}{Nediyana Daskalova}, \bibinfo{person}{Jeff Huang}, {and}
  \bibinfo{person}{James Hays}.} \bibinfo{year}{2016}\natexlab{}.
\newblock \showarticletitle{Webgazer: Scalable webcam ` tracking using user
  interactions}. In \bibinfo{booktitle}{\emph{Proceedings of the Twenty-Fifth
  International Joint Conference on Artificial Intelligence-IJCAI 2016}}.
\newblock


\bibitem[\protect\citeauthoryear{Peters and Itti}{Peters and Itti}{2007}]%
        {peters2007beyond}
\bibfield{author}{\bibinfo{person}{Robert~J Peters} {and}
  \bibinfo{person}{Laurent Itti}.} \bibinfo{year}{2007}\natexlab{}.
\newblock \showarticletitle{Beyond bottom-up: Incorporating task-dependent
  influences into a computational model of spatial attention}. In
  \bibinfo{booktitle}{\emph{Computer Vision and Pattern Recognition, 2007.
  CVPR'07. IEEE Conference on}}. IEEE, \bibinfo{pages}{1--8}.
\newblock


\bibitem[\protect\citeauthoryear{Pfeuffer, Vidal, Turner, Bulling, and
  Gellersen}{Pfeuffer et~al\mbox{.}}{2013}]%
        {pfeuffer2013pursuit}
\bibfield{author}{\bibinfo{person}{Ken Pfeuffer}, \bibinfo{person}{Melodie
  Vidal}, \bibinfo{person}{Jayson Turner}, \bibinfo{person}{Andreas Bulling},
  {and} \bibinfo{person}{Hans Gellersen}.} \bibinfo{year}{2013}\natexlab{}.
\newblock \showarticletitle{Pursuit calibration: Making gaze calibration less
  tedious and more flexible}. In \bibinfo{booktitle}{\emph{Proceedings of the
  26th annual ACM symposium on User interface software and technology}}. ACM,
  \bibinfo{pages}{261--270}.
\newblock


\bibitem[\protect\citeauthoryear{Rizzolatti, Riggio, Dascola, and
  Umilt{\'a}}{Rizzolatti et~al\mbox{.}}{1987}]%
        {rizzolatti1987reorienting}
\bibfield{author}{\bibinfo{person}{Giacomo Rizzolatti}, \bibinfo{person}{Lucia
  Riggio}, \bibinfo{person}{Isabella Dascola}, {and} \bibinfo{person}{Carlo
  Umilt{\'a}}.} \bibinfo{year}{1987}\natexlab{}.
\newblock \showarticletitle{Reorienting attention across the horizontal and
  vertical meridians: evidence in favor of a premotor theory of attention}.
\newblock \bibinfo{journal}{\emph{Neuropsychologia}} \bibinfo{volume}{25},
  \bibinfo{number}{1} (\bibinfo{year}{1987}), \bibinfo{pages}{31--40}.
\newblock


\bibitem[\protect\citeauthoryear{Salvucci and Goldberg}{Salvucci and
  Goldberg}{2000}]%
        {salvucci2000identifying}
\bibfield{author}{\bibinfo{person}{Dario~D Salvucci} {and}
  \bibinfo{person}{Joseph~H Goldberg}.} \bibinfo{year}{2000}\natexlab{}.
\newblock \showarticletitle{Identifying fixations and saccades in eye-tracking
  protocols}. In \bibinfo{booktitle}{\emph{Proceedings of the 2000 symposium on
  Eye tracking research \& applications}}. ACM, \bibinfo{pages}{71--78}.
\newblock


\bibitem[\protect\citeauthoryear{Schaefer, McPhail, and Warren}{Schaefer
  et~al\mbox{.}}{2006}]%
        {schaefer2006image}
\bibfield{author}{\bibinfo{person}{Scott Schaefer}, \bibinfo{person}{Travis
  McPhail}, {and} \bibinfo{person}{Joe Warren}.}
  \bibinfo{year}{2006}\natexlab{}.
\newblock \showarticletitle{Image deformation using moving least squares}. In
  \bibinfo{booktitle}{\emph{ACM transactions on graphics (TOG)}},
  Vol.~\bibinfo{volume}{25}. ACM, \bibinfo{pages}{533--540}.
\newblock


\bibitem[\protect\citeauthoryear{Schiffman}{Schiffman}{1990}]%
        {schiffman1990sensation}
\bibfield{author}{\bibinfo{person}{Harvey~Richard Schiffman}.}
  \bibinfo{year}{1990}\natexlab{}.
\newblock \bibinfo{booktitle}{\emph{Sensation and perception: An integrated
  approach}}.
\newblock \bibinfo{publisher}{Oxford, England: John Wiley \& Sons}.
\newblock


\bibitem[\protect\citeauthoryear{Smit and Van~Gisbergen}{Smit and
  Van~Gisbergen}{1990}]%
        {smit1990analysis}
\bibfield{author}{\bibinfo{person}{AC Smit} {and} \bibinfo{person}{JAM
  Van~Gisbergen}.} \bibinfo{year}{1990}\natexlab{}.
\newblock \showarticletitle{An analysis of curvature in fast and slow human
  saccades}.
\newblock \bibinfo{journal}{\emph{Experimental Brain Research}}
  \bibinfo{volume}{81}, \bibinfo{number}{2} (\bibinfo{year}{1990}),
  \bibinfo{pages}{335--345}.
\newblock


\bibitem[\protect\citeauthoryear{{\v{S}}pakov and Gizatdinova}{{\v{S}}pakov and
  Gizatdinova}{2014}]%
        {vspakov2014real}
\bibfield{author}{\bibinfo{person}{Oleg {\v{S}}pakov} {and}
  \bibinfo{person}{Yulia Gizatdinova}.} \bibinfo{year}{2014}\natexlab{}.
\newblock \showarticletitle{Real-time hidden gaze point correction}. In
  \bibinfo{booktitle}{\emph{Proceedings of the symposium on eye tracking
  research and applications}}. ACM, \bibinfo{pages}{291--294}.
\newblock


\bibitem[\protect\citeauthoryear{Sugano and Bulling}{Sugano and
  Bulling}{2015}]%
        {sugano2015self}
\bibfield{author}{\bibinfo{person}{Yusuke Sugano} {and}
  \bibinfo{person}{Andreas Bulling}.} \bibinfo{year}{2015}\natexlab{}.
\newblock \showarticletitle{Self-calibrating head-mounted eye trackers using
  egocentric visual saliency}. In \bibinfo{booktitle}{\emph{Proceedings of the
  28th Annual ACM Symposium on User Interface Software \& Technology}}. ACM,
  \bibinfo{pages}{363--372}.
\newblock


\bibitem[\protect\citeauthoryear{Sugano, Matsushita, and Sato}{Sugano
  et~al\mbox{.}}{2013}]%
        {sugano2013appearance}
\bibfield{author}{\bibinfo{person}{Yusuke Sugano}, \bibinfo{person}{Yasuyuki
  Matsushita}, {and} \bibinfo{person}{Yoichi Sato}.}
  \bibinfo{year}{2013}\natexlab{}.
\newblock \showarticletitle{Appearance-based gaze estimation using visual
  saliency}.
\newblock \bibinfo{journal}{\emph{IEEE transactions on pattern analysis and
  machine intelligence}} \bibinfo{volume}{35}, \bibinfo{number}{2}
  (\bibinfo{year}{2013}), \bibinfo{pages}{329--341}.
\newblock


\bibitem[\protect\citeauthoryear{Sugano, Matsushita, Sato, and Koike}{Sugano
  et~al\mbox{.}}{2015}]%
        {sugano2015appearance}
\bibfield{author}{\bibinfo{person}{Yusuke Sugano}, \bibinfo{person}{Yasuyuki
  Matsushita}, \bibinfo{person}{Yoichi Sato}, {and} \bibinfo{person}{Hideki
  Koike}.} \bibinfo{year}{2015}\natexlab{}.
\newblock \showarticletitle{Appearance-based gaze estimation with online
  calibration from mouse operations}.
\newblock \bibinfo{journal}{\emph{IEEE Transactions on Human-Machine Systems}}
  \bibinfo{volume}{45}, \bibinfo{number}{6} (\bibinfo{year}{2015}),
  \bibinfo{pages}{750--760}.
\newblock


\bibitem[\protect\citeauthoryear{Tipper, Howard, and Jackson}{Tipper
  et~al\mbox{.}}{1997}]%
        {tipper1997selective}
\bibfield{author}{\bibinfo{person}{Steven~P Tipper}, \bibinfo{person}{Louise~A
  Howard}, {and} \bibinfo{person}{Stephen~R Jackson}.}
  \bibinfo{year}{1997}\natexlab{}.
\newblock \showarticletitle{Selective reaching to grasp: Evidence for
  distractor interference effects}.
\newblock \bibinfo{journal}{\emph{Visual cognition}} \bibinfo{volume}{4},
  \bibinfo{number}{1} (\bibinfo{year}{1997}), \bibinfo{pages}{1--38}.
\newblock


\bibitem[\protect\citeauthoryear{Tripathi and Guenter}{Tripathi and
  Guenter}{2017}]%
        {tripathi2017statistical}
\bibfield{author}{\bibinfo{person}{Subarna Tripathi} {and}
  \bibinfo{person}{Brian Guenter}.} \bibinfo{year}{2017}\natexlab{}.
\newblock \showarticletitle{A statistical approach to continuous
  self-calibrating eye gaze tracking for head-mounted virtual reality systems}.
  In \bibinfo{booktitle}{\emph{Applications of Computer Vision (WACV), 2017
  IEEE Winter Conference on}}. IEEE, \bibinfo{pages}{862--870}.
\newblock


\bibitem[\protect\citeauthoryear{Tudge, McSorley, Brandt, and Schubert}{Tudge
  et~al\mbox{.}}{2017}]%
        {tudge2017setting}
\bibfield{author}{\bibinfo{person}{Luke Tudge}, \bibinfo{person}{Eugene
  McSorley}, \bibinfo{person}{Stephan~A Brandt}, {and} \bibinfo{person}{Torsten
  Schubert}.} \bibinfo{year}{2017}\natexlab{}.
\newblock \showarticletitle{Setting things straight: A comparison of measures
  of saccade trajectory deviation}.
\newblock \bibinfo{journal}{\emph{Behavior research methods}}
  \bibinfo{volume}{49}, \bibinfo{number}{6} (\bibinfo{year}{2017}),
  \bibinfo{pages}{2127--2145}.
\newblock


\bibitem[\protect\citeauthoryear{Van~der Stigchel}{Van~der Stigchel}{2010}]%
        {van2010recent}
\bibfield{author}{\bibinfo{person}{Stefan Van~der Stigchel}.}
  \bibinfo{year}{2010}\natexlab{}.
\newblock \showarticletitle{Recent advances in the study of saccade trajectory
  deviations}.
\newblock \bibinfo{journal}{\emph{Vision research}} \bibinfo{volume}{50},
  \bibinfo{number}{17} (\bibinfo{year}{2010}), \bibinfo{pages}{1619--1627}.
\newblock


\bibitem[\protect\citeauthoryear{Van~der Stigchel, Meeter, and
  Theeuwes}{Van~der Stigchel et~al\mbox{.}}{2006}]%
        {van2006eye}
\bibfield{author}{\bibinfo{person}{Stefan Van~der Stigchel},
  \bibinfo{person}{Martijn Meeter}, {and} \bibinfo{person}{Jan Theeuwes}.}
  \bibinfo{year}{2006}\natexlab{}.
\newblock \showarticletitle{Eye movement trajectories and what they tell us}.
\newblock \bibinfo{journal}{\emph{Neuroscience \& biobehavioral reviews}}
  \bibinfo{volume}{30}, \bibinfo{number}{5} (\bibinfo{year}{2006}),
  \bibinfo{pages}{666--679}.
\newblock


\bibitem[\protect\citeauthoryear{Van~Opstal and Van~Gisbergen}{Van~Opstal and
  Van~Gisbergen}{1987}]%
        {van1987skewness}
\bibfield{author}{\bibinfo{person}{AJ Van~Opstal} {and} \bibinfo{person}{JAM
  Van~Gisbergen}.} \bibinfo{year}{1987}\natexlab{}.
\newblock \showarticletitle{Skewness of saccadic velocity profiles: a unifying
  parameter for normal and slow saccades}.
\newblock \bibinfo{journal}{\emph{Vision research}} \bibinfo{volume}{27},
  \bibinfo{number}{5} (\bibinfo{year}{1987}), \bibinfo{pages}{731--745}.
\newblock


\bibitem[\protect\citeauthoryear{Viviani, Berthoz, and Tracey}{Viviani
  et~al\mbox{.}}{1977}]%
        {viviani1977curvature}
\bibfield{author}{\bibinfo{person}{Paolo Viviani}, \bibinfo{person}{Alain
  Berthoz}, {and} \bibinfo{person}{David Tracey}.}
  \bibinfo{year}{1977}\natexlab{}.
\newblock \showarticletitle{The curvature of oblique saccades}.
\newblock \bibinfo{journal}{\emph{Vision research}} (\bibinfo{year}{1977}).
\newblock


\bibitem[\protect\citeauthoryear{Walker, McSorley, and Haggard}{Walker
  et~al\mbox{.}}{2006}]%
        {walker2006control}
\bibfield{author}{\bibinfo{person}{Robin Walker}, \bibinfo{person}{Eugene
  McSorley}, {and} \bibinfo{person}{Patrick Haggard}.}
  \bibinfo{year}{2006}\natexlab{}.
\newblock \showarticletitle{The control of saccade trajectories: Direction of
  curvature depends on prior knowledge of target location and saccade latency}.
\newblock \bibinfo{journal}{\emph{Perception \& Psychophysics}}
  \bibinfo{volume}{68}, \bibinfo{number}{1} (\bibinfo{year}{2006}),
  \bibinfo{pages}{129--138}.
\newblock


\bibitem[\protect\citeauthoryear{Wang, Wang, and Ji}{Wang
  et~al\mbox{.}}{2016}]%
        {wang2016deep}
\bibfield{author}{\bibinfo{person}{Kang Wang}, \bibinfo{person}{Shen Wang},
  {and} \bibinfo{person}{Qiang Ji}.} \bibinfo{year}{2016}\natexlab{}.
\newblock \showarticletitle{Deep eye fixation map learning for calibration-free
  eye gaze tracking}. In \bibinfo{booktitle}{\emph{Proceedings of the Ninth
  Biennial ACM Symposium on Eye Tracking Research \& Applications}}. ACM,
  \bibinfo{pages}{47--55}.
\newblock


\bibitem[\protect\citeauthoryear{Wang, Satel, Trappenberg, and Klein}{Wang
  et~al\mbox{.}}{2011}]%
        {wang2011aftereffects}
\bibfield{author}{\bibinfo{person}{Zhiguo Wang}, \bibinfo{person}{Jason Satel},
  \bibinfo{person}{Thomas~P Trappenberg}, {and} \bibinfo{person}{Raymond~M
  Klein}.} \bibinfo{year}{2011}\natexlab{}.
\newblock \showarticletitle{Aftereffects of saccades explored in a dynamic
  neural field model of the superior colliculus}.
\newblock \bibinfo{journal}{\emph{Journal of Eye Movement Research}}
  \bibinfo{volume}{4}, \bibinfo{number}{2} (\bibinfo{year}{2011}).
\newblock


\bibitem[\protect\citeauthoryear{Xu, Jiang, Wang, Kankanhalli, and Zhao}{Xu
  et~al\mbox{.}}{2014}]%
        {xu2014predicting}
\bibfield{author}{\bibinfo{person}{Juan Xu}, \bibinfo{person}{Ming Jiang},
  \bibinfo{person}{Shuo Wang}, \bibinfo{person}{Mohan~S Kankanhalli}, {and}
  \bibinfo{person}{Qi Zhao}.} \bibinfo{year}{2014}\natexlab{}.
\newblock \showarticletitle{Predicting human gaze beyond pixels}.
\newblock \bibinfo{journal}{\emph{Journal of vision}} \bibinfo{volume}{14},
  \bibinfo{number}{1} (\bibinfo{year}{2014}), \bibinfo{pages}{28--28}.
\newblock


\bibitem[\protect\citeauthoryear{Yarbus}{Yarbus}{1967}]%
        {yarbus1967eye}
\bibfield{author}{\bibinfo{person}{AL Yarbus}.}
  \bibinfo{year}{1967}\natexlab{}.
\newblock \showarticletitle{Eye movements and vision. 1967}.
\newblock \bibinfo{journal}{\emph{New York}} (\bibinfo{year}{1967}).
\newblock


\bibitem[\protect\citeauthoryear{Zhang, Huang, Sugano, and Bulling}{Zhang
  et~al\mbox{.}}{2018}]%
        {zhang2018training}
\bibfield{author}{\bibinfo{person}{Xucong Zhang},
  \bibinfo{person}{Michael~Xuelin Huang}, \bibinfo{person}{Yusuke Sugano},
  {and} \bibinfo{person}{Andreas Bulling}.} \bibinfo{year}{2018}\natexlab{}.
\newblock \showarticletitle{Training Person-Specific Gaze Estimators from User
  Interactions with Multiple Devices}. In \bibinfo{booktitle}{\emph{Proceedings
  of the 2018 CHI Conference on Human Factors in Computing Systems}}. ACM,
  \bibinfo{pages}{624}.
\newblock


\end{thebibliography}
